\documentclass[a4paper,11pt]{article}
\pdfoutput=1 

\usepackage{jcappub} 

\usepackage[T1]{fontenc} 

\usepackage{bm,graphicx,dcolumn,epstopdf,epsf, latexsym,mathbbol, amssymb,amsmath,color,slashed, mathrsfs,mathcomp,simplewick}

\usepackage{graphicx}
\usepackage{dcolumn}
\usepackage{bm}
\usepackage{latexsym}
\usepackage{amsfonts}
\usepackage{amssymb}
\usepackage{amsmath}
\usepackage{color,epstopdf,epsf, latexsym,mathbbol,slashed, mathrsfs,mathcomp,simplewick}
\usepackage[center]{subfigure}
\usepackage{multirow}
\usepackage{makecell}

\renewcommand{\theequation}{1.\arabic{equation}} 
 \newcommand{\bq}{\begin{equation}}
 \newcommand{\eq}{\end{equation}}
 \newcommand{\bqn}{\begin{eqnarray}}
 \newcommand{\eqn}{\end{eqnarray}}
 \newcommand{\nb}{\nonumber}
 \newcommand{\lb}{\label}

\title{Schwinger Pair Production by Electric Field Coupled to Inflaton}

\author{ Jia-Jia Geng $^{a, b}$}
\emailAdd{Jiajia$\_$Geng@baylor.edu}

\author{Bao-Fei Li $^{a, b}$}
\emailAdd{Bao-Fei$\_$Li@baylor.edu}

\author{Jiro Soda$^{c}$}
\emailAdd{jiro@phys.sci.kobe-u.ac.jp}

\author{Anzhong Wang $^{a, b}$}
\emailAdd{Anzhong$\_$Wang@baylor.edu}

\author{Qiang Wu $^{a}$}
\emailAdd{wuq@zjut.edu.cn}

\author{Tao Zhu $^{a, b}$}
\emailAdd{Tao$\_$Zhu@baylor.edu}

\affiliation{ ${}^{a}$ Institute for Advanced Physics $\&$ Mathematics, Zhejiang University of Technology, Hangzhou, 310032, China\\
${}^{b}$ GCAP-CASPER, Physics Department, Baylor University, Waco, TX 76798-7316, USA \\
$^{c}$ Department of Physics, Kobe University, Kobe 657-8501, Japan}

\abstract{
We analytically investigate the Schwinger pair production in  the de Sitter background by using {\em the uniform asymptotic approximation method}, 
and show that the equation of motion in general has two turning points, and the nature of these points could be single, double, real or complex, 
depending on the choice of the free parameters involved in the theory.  Different natures of these points lead to different electric currents. In particular, 
when $\beta \equiv m^2/H^2-9/4$ is positive, both turning points are complex,  and the electric current due to the Schwinger process is highly suppressed, 
where $m$ and $H$ denote, respectively, the mass of the particle and the Hubble parameter. For the turning points to be real, it is necessary to have 
$\beta < 0$, and  the more negative of $\beta$,  the easier to produce particles.  In addition, when $\beta < 0$, we also study the particle production 
when the electric field $E$ is very weak. We find  that the electric current  in this case is proportional to $E^{1/2 - \sqrt{|\beta|}}$, which is strongly enhanced 
in the weak electric field limit when  $m < \sqrt{2} H$. 
}

\begin{document}

\maketitle

\section{Introduction}
\renewcommand{\theequation}{1.\arabic{equation}} \setcounter{equation}{0}
\label{sc:basic}

It is well known that the strong electric field causes particle pair production named the Schwinger effect~\cite{Schwinger:1951nm}. This occurs when the 
electrostatic energy $qEL$ exceeds the twice of the rest mass $m$ of charged particles, namely, $qEL>2m$. Here, $q$ is the charge of particles, $E$ is the electric field, 
and $L$ is a typical length scale. As for the length scale $L$, we usually take the Compton wave length of the particle $1/m$. Hence, the condition for Schwinger 
pair production becomes $qE>2m^2$.  Since the Schwinger effect in (1+1)-dimensions can be interpretted as the nucleation process of pairs of particles.
it has been studied as a useful model of the bubble nucleation in (3+1)-dimensions. 
For example, observer dependence of bubble nucleation process has been explored using Schwinger pair production process as a model~\cite{Garriga:2012qp,Garriga:2013pga}.

It is also intriguing to consider the Schwinger pair production in de Sitter space from the cosmological point of view. 
 In the cosmological Schwinger effect, there is another length scale, that is, the Hubble parameter. 
 In particular, when the mass of a charged particle is sufficiently small, the particle picture is obscure. 
 Therefore, it is interesting to investigate Schwinger pair production in the expanding universe. 
There are several works studying Schwinger pair production in the expanding (1+1)-dimensional universe~\cite{Garriga:1994bm, Kim:2008xv, Kim:2010cb}. 
Especially, in \cite{Frob:2014zka}, an interesting phenomena called infrared hyperconductivity was found. 
This occurs for light fields. There, the current created by pair production increases as the electric field decreases in contradiction to
the intuition.  
Recently, the analysis has been extended to 4-dimensional expanding universe~\cite{Kobayashi:2014zza,FNPT15,Hayashinaka:2016qqn,Hayashinaka:2016dnt,SS17}. 
However, in these works, the electric field in terms of energy density is decaying.
On the other hand, for example, the energy density of the electric field in anisotropic inflation does not decay~\cite{Watanabe:2009ct,Soda:2012zm,Maleknejad:2012fw}.
Actually, when we consider the generation of magnetic fields during inflation~\cite{Ratra:1991bn}, this kind of non-decaying electric field is generic.
Indeed, if an inflaton couples to the electric field through the  gauge kinetic function, the electric field remains during inflation. 
Moreover, the scale invariant growth is an attractor solution~\cite{Kanno:2009ei}.   
Hence, it is interesting to consider the Schwinger effect in this set up.
It would be nice to know the electric current and energy momentum tensor due to the Schwinger pair production 
and clarify the backreaction to anisotropic inflation, in particular, anisotropic inflation with a charged inflaton~\cite{Emami:2010rm}. 

In this paper, as a first step, we analytically study the Schwinger pair production using {\em the uniform asymptotic approximation method}
\cite{zhu_constructing_2014,zhu_gravitational_2014,zhu_inflationary_2014,zhu_high-order_2016},  and 
calculate particle production rate due to the persistent electric field
in  the (3+1)-dimensional  de Sitter space. We focus on the parameter region where the hyperconductivity is expected to occur, 
namely, we consider the weak electric field and the light scalar field. 
We find the particle production rate grows as $ E^{1/2-\sqrt{|\beta|}}$ for $\beta \equiv m^2/H^2 - 9/4 < 0$.
This remarkable behavior comes from the infrared behavior of the light fields in de Sitter space.

The organization of the rest parts of the paper is as follows.
In section II, we introduce the basic setup. We consider the persistent electric field sustained by an inflaton field.
On this background, we consider  a charged scalar field.  
In section III,  we apply the uniform asymptotic approximation method to the Schwinger effect in the (3+1)-dimensional de Sitter space
and obtain the Bogoliubov coefficients.
In section IV, we evaluate the particle production rate in de Sitter space in the presence of the electric field.
The final section is devoted to the conclusion. We gather technical parts in the four appendixes. In particular, in Appendix D we apply the uniform asymptotic approximation method
to study the pair production of particles in (1+1)-dimensional de Sitter space, a case in which the exact mode function is known \cite{Garriga:1994bm}, so we can  compare
the approximate solution obtained by the uniform asymptotic approximation method with the exact one. From Fig. \ref{2dcase}, it can be  seen that  
 the approximate solution traces  the exact one extremely well.

\section{Electric field during anisotropic Inflation}
\renewcommand{\theequation}{2.\arabic{equation}} \setcounter{equation}{0}

In this section, we gives a basic setup we are going to investigate.
The point is that the gauge field has an attractor solution in the inflationary background.

Let us consider the following action for the gravitational field, the inflaton field $\varphi$ and the electro-magnetic vector field $A_\mu$ coupled with $\varphi$:
\begin{eqnarray}
S=\int d^4x\sqrt{-g}\left[~\frac{M_p^2}{2}R
-\frac{1}{2}\left(\partial_\mu\varphi\right)\left(\partial^{\mu}\varphi\right)
-V(\varphi)-\frac{1}{4} f^2 (\varphi) F_{\mu\nu}F^{\mu\nu}
~\right] \ ,
\label{action1}
\end{eqnarray}
where $M_p$ is the reduced Plack mass, $g$ is the determinant of the metric, $R$ is the Ricci scalar, $V(\varphi)$ is the inflaton potential, $f(\varphi)$ is the coupling 
function of the inflaton field to the vector one, respectively. The field strength of the vector field is defined by $F_{\mu\nu}=\partial_\mu A_\nu -\partial_\nu A_\mu$. If
 $f(\varphi)$ is a constant, nothing happens for the gauge field. However, if $f(\varphi)$ rapidly changes, the gauge field will be sustained during inflation. In terms 
 of the time dependence, the critical case corresponds to 
$$
f(\varphi) \propto \frac{1}{a^2} \ .
$$
Here $a$ denotes the scale factor of the de-Sitter background and is introduced in (\ref{FRW}) below. Actually, this is known as an attractor even for supercritical cases 
when we take into account the backreaction. This attractor solution has been named anisotropic inflation~\cite{Watanabe:2009ct}.

We can look at anisotropic inflation from the background and derturbative point of view. 
First, we  ignore the tiny  anisotropy of expansion.
Hence, we can consider a fixed isotropic de Sitter background
\begin{eqnarray}
ds^2 = a^2 (\eta) \left( -d\eta^2 + dx^2 + dy^2 + dz^2 \right) \ ,
\label{FRW}
\end{eqnarray}
where $a = 1/(-H\eta)$ and we used the conformal time $\eta$.
Then, we put the Maxwell field on this background as a test field
\begin{eqnarray}
S=\int d^4x\sqrt{-g}\left[-\frac{1}{4} f^2 (\varphi) F_{\mu\nu}F^{\mu\nu}
~\right] \ ,
\label{action1}
\end{eqnarray}
where we assumed a general coupling function.
Now, we take the homogeneous fields of the form
\begin{eqnarray}
\lb{amu1}
A_\mu=(~0,~v(\eta),~0,~0~)
\end{eqnarray}
The equation of motion $\nabla_\mu \left( f^2 F^{\mu\nu} \right) =0$
can be solved as
\begin{eqnarray}
\lb{amu2}
    v(\eta) = E \int d\eta f^{-2}
            = \frac{E}{(2\alpha -1) H^{2\alpha}} (-\eta)^{-2\alpha +1}  \ .
\end{eqnarray}
where we  assumed $ f = a^{-\alpha}$. The functional form of $f$, that is, the parameter $\alpha$
should be determined by the model $f(\varphi)$ we are considering. 
The trivial case $\alpha =0$ corresponds to the trivial function $f=1$ where the gage field rapidly decays as expected.
In anisotropic inflation, we know $\alpha =2$ is an attractor~\cite{Kanno:2009ei}.
In this case, the gauge field grows. 
What we should investigate is Schwinger effect in this background.

We gave a perturbative model for anisotropic inflation.
Now, we regard these configurations as the background and 
 consider a test charged scalar field \footnote{In this paper, we choose units so that the electric charge $e = 1$. To restore the unit, we can simply replace $E$ by $eE$, as one can see from
 Eqs.(\ref{amu1})-(\ref{amu3}).}
\begin{eqnarray}
\lb{amu3}
  S = \int d^4 x \sqrt{-g} \left[
      - g^{\mu\nu}\left( \partial_\mu + i e A_\mu \right) \psi^*
      \left( \partial_\nu - i e A_\nu \right) \psi -m^2 \psi^* \psi
       \right] \ .
\end{eqnarray}
The equation of motion can be deduced as
\begin{eqnarray}
  \frac{1}{\sqrt{-g}} i\partial_\mu \left[ \sqrt{-g} g^{\mu\nu}
  \left( i\partial_\nu + A_\nu \right) \psi \right]
  + g^{\mu\nu} A_\mu \left( i\partial_\nu + A_\nu \right) \psi
  + m^2 \psi = 0 \ .
\end{eqnarray}
In the background, we obtain
\begin{eqnarray}
  \psi'' + 2 \frac{a'}{a} \psi' - \partial_a^2 \psi
  + \left( i\partial_x + v \right)^2 \psi + m^2 a^2 \psi =0 \ .
\end{eqnarray}
By defining a new field $\phi (\eta) e^{i p_a x^a +ikx} = a\psi$, we have
\begin{eqnarray}
   \phi'' + \left[ p_a^2
   +\left( k -\frac{E}{(2\alpha -1) H^{2\alpha}} (-\eta)^{-2\alpha +1}\right)^2
   + \frac{m^2}{H^2 \eta^2} -\frac{2}{\eta^2} \right] \phi =0 \ .
\end{eqnarray}

Note that  $\alpha =1$ is a special case for which we can solve the wave equation exactly.
The result is quite similar to (1+1)-dimensional problem and analyzed in 
\cite{Kobayashi:2014zza,Hayashinaka:2016qqn,Hayashinaka:2016dnt}.
However, what we are interested in is the case $\alpha =2$ which stems from anisotropic inflation.
In this case, the equation reads
\begin{eqnarray}
   \phi'' + \left[ p_a^2
   +\left( k -\frac{E}{3 H^{4}} (-\eta)^{-3}\right)^2
   + \frac{m^2}{H^2 \eta^2} -\frac{2}{\eta^2} \right] \phi =0 \ .
\end{eqnarray}
In the following sections, we will focus on the Schwinger pair production process with this setup.
Apparently, the equation is not the Fuchs type. Hence, we need to resort to approximation or numerical method.
We will do both in subsequent sections.

\section{ Uniform asymptotic approximation}
\renewcommand{\theequation}{3.\arabic{equation}} \setcounter{equation}{0}

In this section, we are going to apply the {\em uniform asymptotic approximation} developed in a series of  papers 
\cite{zhu_constructing_2014,zhu_gravitational_2014,zhu_inflationary_2014,zhu_high-order_2016} to study the above equation. For this purpose, it is convenient to write the above equation in the form
\bqn\lb{eom}
\frac{d^2 \phi_{k_\pm}}{dy^2}+\left(d+\frac{\tilde \beta}{y^2}+\frac{b_{\pm}}{y^3}+\frac{c}{y^6}\right)\phi_{k_\pm}=0,
\eqn
where $k_+$($k_-$) denotes positive(negative) k, $y \equiv- |k| \eta$,  $\phi_{k_\pm}(y)=\phi(\eta)$, and the parameters $d, \tilde \beta, b_\pm, c$ are defined by
\bqn
\lb{parameters}
\tilde \beta \equiv \frac{m^2}{H^2}-2,\;\;\;\; b_\pm\equiv  \mp\frac{2 k^2 E}{3H^4}, \;\;\; c\equiv \frac{b_\pm^2}{4}=\left(\frac{k^2E}{3H^4}\right)^2, \;\;\;\; d\equiv \frac{p_a^2}{k^2}+1.
\eqn
Here $b_+$($b_-$) corresponds to $k_+$($k_-$). Note that in de Sitter background, the variable $y$ is in the range $y \in (0^+, +\infty)$. 
Following \cite{zhu_constructing_2014,zhu_gravitational_2014,zhu_inflationary_2014,zhu_high-order_2016}, let us first write Eq.(\ref{eom}) in the standard form
\bqn\lb{eom_uniform}
\frac{d^2 \phi_{k_\pm}}{dy^2}= \{\lambda^2 \hat g(y)+q(y)\} \phi_{k_\pm},
\eqn
where
\bqn\lb{gplusq}
\lambda^2 \hat g(y)+q(y)=- \left(d+\frac{\tilde \beta}{y^2}+\frac{b_\pm}{y^3}+\frac{c}{y^6}\right),
\eqn
in which $\lambda$ is supposed to be a large parameter and will be used to trace the approximate orders of the uniform asymptotic approximation. At the end,   we can set $\lambda=1$ for simplicity. 
The functions  $g(y)$ and $q(y)$ are arbitrary. The reason to introduce two of them, instead of only one,  is to use one of them to minimize the errors, as to be shown below. When $b_\pm=0$, 
 Eq.(\ref{eom_uniform}) has exact solution. Thus, in the following we are going to concentrate ourselves only on the cases where  $b_\pm  \neq 0$.

\subsection{Errors of the approximation near the pole}

From the expression (\ref{gplusq}), one can see that the functions $\lambda^2 \hat g(y)$ in general  has a pole at $y=0$, which is of order higher than $2$. Following \cite{zhu_inflationary_2014}, in order to determine the two functions
 $\lambda^2 \hat g(y)$ and $q(y)$, we first analyze the behavior of the corresponding error control function of the uniform asymptotic approximation near this pole. As shown in \cite{zhu_inflationary_2014}, the error control function 
near the pole $y=0^+$ is expressed as \footnote{The detailed derivation of the approximation and the corresponding error control functuon can be found in \cite{zhu_inflationary_2014}.},
\bqn
\mathscr{F}(y) =\int^y \left[\frac{5}{16} \frac{\hat g'^2(y)}{\hat g^{5/2}(y)} - \frac{1}{4} \frac{\hat g''(y)}{\hat g^{3/2}(y)}-\frac{q(y)}{\hat g^{1/2}(y)}\right]dy.
\eqn
A crucial step in the application of the uniform asymptotic approximation is to guarantee that the error control function $\mathscr{F}(y)$ is finite near pole $y = 0^+$. Otherwise the approximate solution of Eq.(\ref{eom_uniform}) constructed by
 the uniform asymptotic approximation will blow up when $y \rightarrow 0^+$. For this purpose, following \cite{zhu_inflationary_2014}, we expand $\lambda^2 \hat g(y)$ and $q(y)$  in the Laurent series around $y=0$ as
\bqn
\lb{expand function}
\lambda^2 \hat g(y)=\frac{1}{y^i}\sum_{s=0}^\infty g_s y^{s},\;q(y)=\frac{1}{y^j} \sum_{s=0}^\infty q_s y^s,
\eqn
where $i$ and $j$ represent the order of the pole at $y=0$ for $\lambda^2 \hat g(y)$ and $q(y)$,  respectively. Then,  to the  leading order, we find that 
\bqn
\lb{error}
\mathscr{F}(y)  
\simeq - \frac{1}{g_0^{1/2}} \int^y \left[ \left(\frac{5 i^2}{16} -\frac{i(i+1)}{4}\right)y^{\frac{i}{2}-2}-q_0y^{\frac{i}{2}-j} \right]dy.
\eqn 
Thus,  to keep the error control function $\mathscr{F}(y)$ finite near $ y = 0^+$, we must require,
\bq
\lb{cds}
i \ge 2, \;\;\; j \le \frac{i}{2} + 1.
\eq
However, considering Eq.(\ref{gplusq}) we find that  the choices of $i$ and $j$ also depend on the values of $b_\pm$. In particular, when  $b_\pm \neq 0$, 
we have $i=6$ and $j\le 4$. When  $b_\pm = 0$, we have $i  = 2$. In this case,  to keep the error control function $\mathscr{F}(y)$ be finite nearby  the pole, we must also choose $j = 2$ and $q_0 = -1/4$, that is, 
\bqn
\lb{cdq}
q(y) = - \frac{1}{ 4y^2}.
\eqn
 In addition, even in the case  $b_\pm \neq 0$, if $|b_\pm| \ll 1$ (As to be shown below, this exactly corresponds to the weak electric limit.), there always exists a region 
 $ 1 \gg y > \left|b_\pm^2/(4\tilde{\beta})\right|^{1/4}$, in which the term $\tilde{\beta}/y^2$ in Eq.(\ref{gplusq}) dominates,
and to minimize the error control function $\mathscr{F}(y)$, Eq.(\ref{cdq}) still represents a good choice. Yet, even when the external electric field 
 is large,  the error control function for the choice of Eq.(\ref{cdq}) still remains finite.  So,  in this paper we shall adopt Eq.(\ref{cdq}) 
 for all the cases. With this choice, as to be shown below,   the analytical solution traces   the numerical (exact) one extremely well.  Then, we obtain 
\bqn\lb{general_g}
\lambda^2 \hat g(y) = - \left(d+\frac{\beta}{y^2}+\frac{b_\pm}{y^3}+\frac{c}{y^6}\right).
\eqn
with $\beta\equiv \tilde \beta-1/4 = m^2/H^2-9/4$. 

\subsection{Nature of the turning points}
\lb{turning_points}

Except that $\lambda^2 \hat g(y)$ has a pole at $y=0^+$, it may also have zeros in the range $y \in (0^+,+\infty)$, which are called {\em turning points}. As shown in \cite{zhu_inflationary_2014}, the 
corresponding approximate solution of (\ref{eom}) in the uniform asymptotic approximation is very sensitive to the nature of these turning points, which are  the roots of the equation, 
\bqn
\lb{roots}
 d+\frac{\beta}{y^2}+\frac{b_\pm}{y^3}+\frac{c}{y^6} = 0.
\eqn
From the definition of parameters $\beta, b_\pm, c, d$, we find that
\bqn
\lambda^2 \hat g(y) =
\begin{cases}
-d <0, &  \;\; y\to +\infty,\\ 
- \frac{c}{y^6} <0, & \;\;\; y\to 0^+.
\end{cases}
\eqn
In principle, Eq. (\ref{roots}) can have six roots. Since $d >0$, in both limits $y \to 0^{+}$ and $y \to +\infty$, the function $\lambda^2 \hat g(y)$ is negative. This implies that $\lambda^2 \hat g(y)=0$ can only have even numbers of roots, i.e., it can only have six, four, two, or zero roots. We note that these roots can be real, multiple, or even complex conjugated. In order to see explicitly how many roots, we can count the extreme points of $d y^6 + \beta y^4 + b_{\pm} y^3 + c$. Taking the derivative of $d y^6 + \beta y^4 + b_{\pm} y^3 + c$ we find
\bqn\lb{extreme}
6 d y^3 + 4 \beta y + 3 b_{\pm}=0.
\eqn
Unlike the roots of (\ref{roots}), the extreme points obtained from (\ref{extreme}) must be real and lie in the range $y \in (0, +\infty)$. It can be shown that the three extreme points must satisfied the relation $y_{m1}+y_{m2} + y_{m3}=0$. This indicates that one of them must lie in the negative side of the $y$-axix. Thus, in the positive side, $\lambda^2 \hat g(y)$  can at most have two extreme points, which implies that  it can at most  have two turning points. In Fig.~\ref{gofy_figure_1} we show several representative cases, which tells us that this is indeed the case.

\begin{figure}
{\label{gofy_figure_1}
\includegraphics[width=12.1cm]{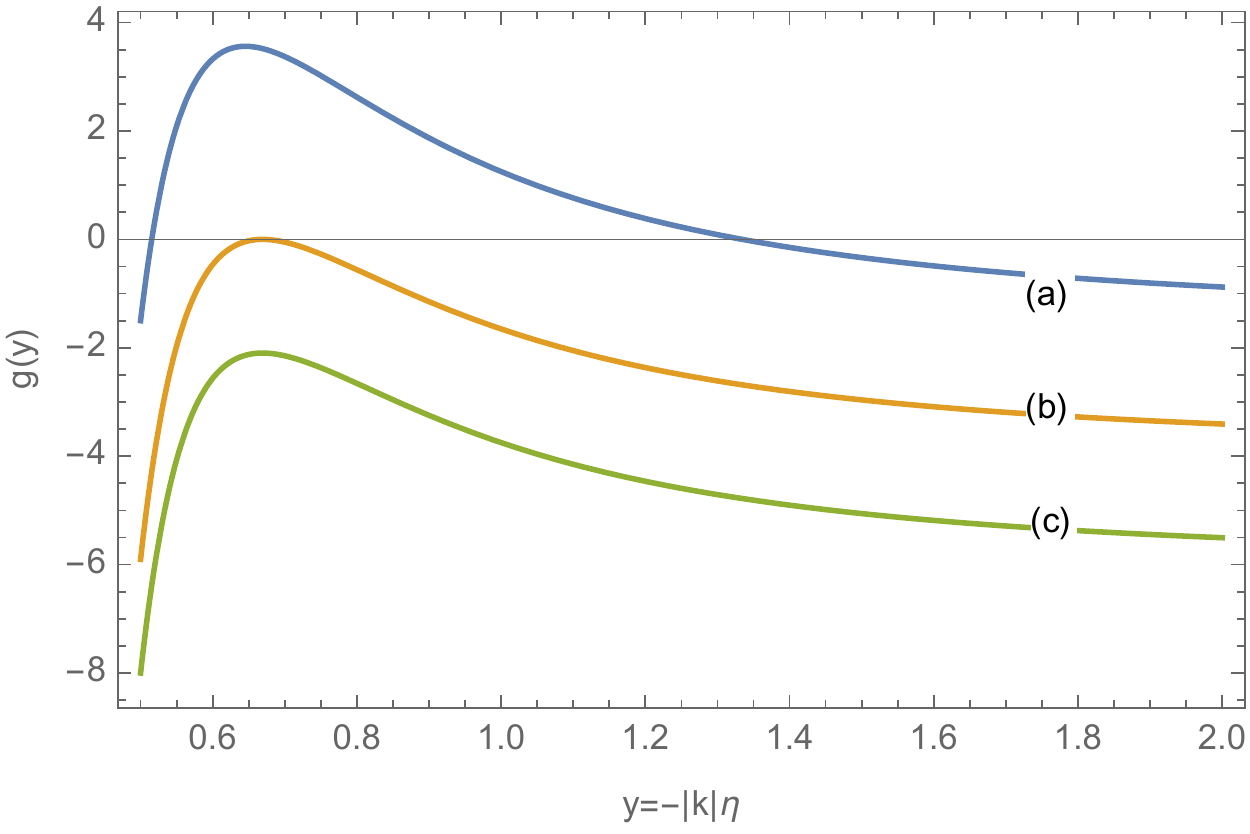}}
\caption{The three different behaviors of the function $\lambda^2 \hat g(y)$ in Eq.(\ref{general_g}). For each case, $\lambda^2 \hat g(y)$ has different types of turning points (zeros). 
Case (a, b, c) correspond, respectively,  to  two real and single, one real and double, and two complex conjugated turning points. When $k>0$, the value of parameters used for each case are: 
(a) $\beta=-2 (m^2/H^2=1/4)$, $b=-1 (k^2E/H^4=3/2)$, $c=1/4$, $d=3/2 (p_a^2/k^2=1/2)$; (b) $\beta=-3/2\; (m^2/H^2=3/4)$, $b=-1\; (k^2E/H^4=3/2)$, $c=1/4$, $d=3.9027 \; 
(p_a^2/k^2=2.9027)$; (c) $\beta=-3/2\; (m^2/H^2=3/4)$, $b=-1\; (k^2E/H^4=3/2)$, $c=1/4$, $d=6 \; (p_a^2/k^2=5)$. For $k<0$, in order to get these three different cases, one must
choose  the parameters as: (a) $\beta=-2 (m^2/H^2=1/4)$, $b=0.15 (k^2E/H^4=0.225)$, $c=0.0375$, $d=1.05 (p_a^2/k^2=0.05)$; (b) $\beta=-3/2\; (m^2/H^2=3/4)$, $b=0.15\;
 (k^2E/H^4=0.225)$, $c=0.0375$, $d=2.714 \; (p_a^2/k^2=1.714)$; (c) $\beta=-3/2\; (m^2/H^2=3/4)$, $b=0.15\; (k^2E/H^4=0.225)$, $c=0.0375$, $d=6 \; (p_a^2/k^2=5)$.}
\label{gofy_figure_1}
\end{figure}

To study these roots further, let us first write  the function $\lambda^2 \hat g(y)$ in the form
\bqn
\lb{3b}
\lambda^2 \hat g(y) = - \left(\frac{b_\pm}{2y^3}+1\right)^2 - \frac{p_a^2}{k^2} - \frac{\beta}{y^2}.
\eqn
To find out the roots of $\lambda^2 \hat g(y)$ we need to solve the equation, 
\bqn\lb{equation}
 \left(\frac{b_\pm}{2y^3}+1\right)^2 = - \frac{\beta}{y^2}- \frac{p_a^2}{k^2}.
\eqn
An inspection on this equation immediately leads to the conclusion that the function $\lambda^2 \hat g(y)$ can have real turning points only when
\bqn
(i) \; \beta<0 \;\;\; \text{or} \;\;\;  (ii)\; \beta=0= p_a.
\eqn
In general, if the two turning points are  not equal (or a double root) of Eq.(\ref{equation}), each of them must be real and single. We denote these two turning points by $y_1$ and $y_2$. 
When the values of the parameters change, these two turning points can eventually turn to a double or even complex  turning points. In any  case, we  assume that $\text{Re}(y_1)\leq 
\text{Re}(y_2)$, respectively. In the following, let us first  consider two specific cases in which   $y_1$ and $y_2$ can be  derived explicitly.

\subsubsection{$\beta=0$}
In this case, we find that 
\bqn
 \left(\frac{b_\pm}{2y^3}+1\right)^2 = - \frac{p_a^2}{k^2},
\eqn
which leads to
\bqn\lb{y1y2_caseA}
y_1^3 &=& -\frac{b_\pm}{2d} - i \frac{b_\pm}{2d} \frac{p_a}{|k|} ,\;\; y_2^3 = - \frac{b_\pm}{2d} +  i \frac{b_\pm}{2d} \frac{p_a}{|k|},
\eqn
where we assume that $p_a$ is positive. It is easy to see that in this case $y_1$ and $y_2$ are always complex and $y_1 = y_2^*$. Thus in this case there are no real turning points. 

\subsubsection{Weak electric field limit}

In the weak electric field limit, we have
\bqn
|b_\pm|\equiv \frac{2 k^2 E}{3 H^4} \ll 1.
\eqn
In this case we can treat the parameter $b_\pm$ as a small quantity and expand the turning points $y_1$ and $y_2$ in the powers of $b_\pm$. After some calculations  we obtain, for $k>0$,
\bqn\lb{y1y2_CaseC}
y_{1}=\frac{\sqrt{-b_+}}{(-4\beta)^{1/4}}+\frac{b_+}{4\beta}+\mathcal{O}\Big((-b_+)^{3/2}\Big),\;\;\;\; y_{2}=\sqrt{\frac{-\beta}{d}}+\frac{b_+}{2\beta}+\mathcal{O}\Big((-b_+)^2\Big).
\eqn
Correspondingly, for $k<0$ we find
\bqn\lb{y1y2_CaseD}
y_{1}=\frac{\sqrt{b_-}}{(-4\beta)^{1/4}}-\frac{b_-}{4\beta}+\mathcal{O}\Big((b_-)^{3/2}\Big),\;\;\;\; y_{2}=\sqrt{\frac{-\beta}{d}}+\frac{b_-}{2\beta}+\mathcal{O}\Big((b_-)^2\Big).
\eqn
Note that in the above we have assumed that $\beta<0$, and then it is easy to see that $y_1$ and $y_2$ are both real and single.

\subsection{General  solutions with two turning points} 

From the above we can see that   the function  $\lambda^2 \hat g(y)$ at most has  two turning points. In the following, we shall develop the general formulas for the cases with two turning points. 
In the next subsection we consider particle creations first for the general case, and then   the particular cases considered above.  From  \cite{zhu_inflationary_2014}, we find that
 the general approximate solution of Eq.(\ref{eom}) can be expressed in terms of the parabolic cylinder functions $W(\frac{1}{2}\lambda \zeta_0^2,\pm \sqrt{2\lambda} \zeta)$ as
\bqn\lb{solution}
\phi_{k_\pm}(y)&=&\alpha_1 \left(\frac{\zeta^2-\zeta_0^2}{-\hat g(y)}\right)^{\frac{1}{4}} \left[W\left(\frac{1}{2}\lambda \zeta_0^2, \sqrt{2\lambda}\zeta \right)\right]+\beta_1 \left(\frac{\zeta^2-\zeta_0^2}{-\hat g(y)}\right)^{\frac{1}{4}} \left[W\left(\frac{1}{2}\lambda \zeta_0^2, -\sqrt{2\lambda }\zeta \right)\right],\nb\\
\eqn
where $\hat g(y)$ depends on $b_\pm$ through Eq.(\ref{3b})$, \alpha_1$, $\beta_1$ are two integration constants, $\zeta$ is a monotoning function of $y$ which are given in Appendix~\ref{Uniform_App}, and $\zeta_0^2$ is defined as
\bqn
\lb{zeta0} 
\zeta_0^2 = \frac{2}{\pi} \int_{y_1}^{y_2} \sqrt{\hat g(y)} dy.
\eqn

In order to determine the integration constants $\alpha_1$ and $\beta_1$, we need to match the above solution to the initial conditions. In general, we consider the initial time at  $y = - |k|\eta \to +\infty$, for which we have
\bqn
\lambda^2 \hat g(y) \simeq -d.
\eqn
Thus if we consider the Bunch-Davies vacuum in  the limit $y \to +\infty$, then the initial state can be approximately described by
\bqn\lb{initial}
\lim_{y\to +\infty}\phi_{k_\pm}(y) = \frac{e^{ i d^{1/2} y}}{\sqrt{2} d^{1/4}}.
\eqn
In order to match it with the approximate solution we obtained above, let us first consider the asymptotic form of the parabolic cylinder function in the limit $y \to +\infty$ ($\zeta \to +\infty$), which takes the form (see Appendix~\ref{Asy_W})
\bqn
W(\text{\textonehalf}\lambda\zeta_0^2,\sqrt{2\lambda}\zeta)=\left(\frac{ 2 j^2(\sqrt{\lambda}\zeta_0)}{\lambda (\zeta^2-\zeta_0^2)}\right)^{1/4}\cos{\mathfrak D},\;\;\;\; W(\text{\textonehalf}\lambda\zeta_0^2,-\sqrt{2\lambda}\zeta)=\left(\frac{ 2 j^{-2}(\sqrt{\lambda}\zeta_0)}{\lambda (\zeta^2-\zeta_0^2)}\right)^{1/4}\sin{\mathfrak D}.\nb\\
\eqn
Here
\bqn
\mathfrak{D}\equiv \lambda \int_{\text{Re}(y_2)}^{y} \sqrt{-\hat g}dy+\frac{\pi}{4}+\phi\left(\frac{\lambda}{2}\zeta_0^2\right),
\eqn
and
\bqn
\lb{jf}
j(\sqrt{\lambda} \zeta_0)=\sqrt{1+e^{ \pi \lambda \zeta_0^2 }}-e^{\pi \lambda \zeta_0^2/2} \ ,
\label{jform}
\eqn
where $\phi(x)$ is give by Eq.(\ref{phi}), and should not be confused with the  mode function $\phi_k(y)$. With the above asymptotic forms, when $y\to +\infty$ we have
\bqn
\phi_{k_\pm}(y)\simeq \alpha_1 \left(\frac{2 j^2}{-\lambda \hat g}\right)^{1/4}  \cos{\mathfrak{D}}+\beta_1  \left(\frac{2 j^{-2}}{-\lambda \hat g}\right)^{1/4}\sin{\mathfrak{D}},\nb\\
\eqn
here $j\equiv j(\sqrt{\lambda} \zeta_0)$. Matching the above solution with the initial conditions (\ref{initial})  we obtain
\bqn\lb{a1b1}
\alpha_1 = \frac{e^{i\theta}}{\sqrt{2} (2 \lambda j^2)^{1/4}},\;\;\beta_1= \frac{i e^{i\theta}}{\sqrt{2} (2 \lambda j^{-2})^{1/4}},
\eqn
where $\theta$ is an irrelevant phase factor.

\begin{figure}
{\includegraphics[width=8.1cm]{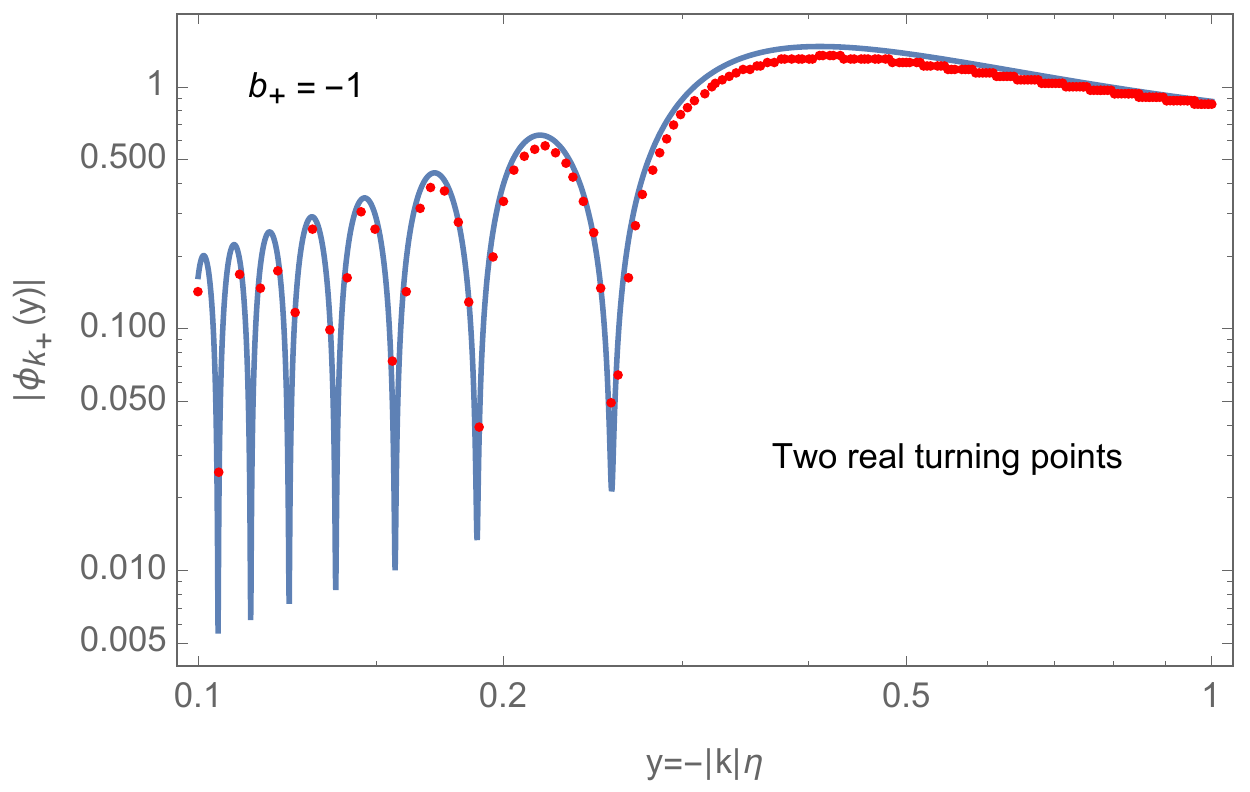}}
{\includegraphics[width=8.1cm]{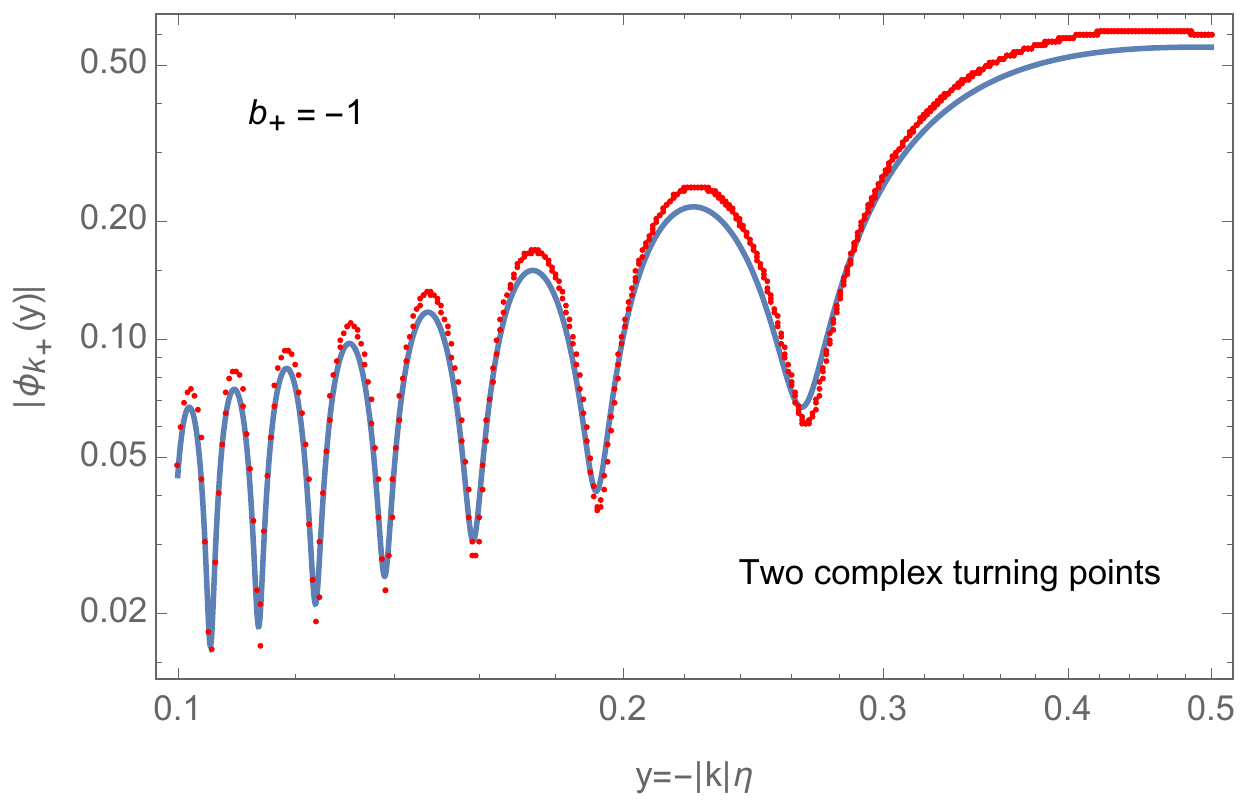}}\\
{\includegraphics[width=8.1cm]{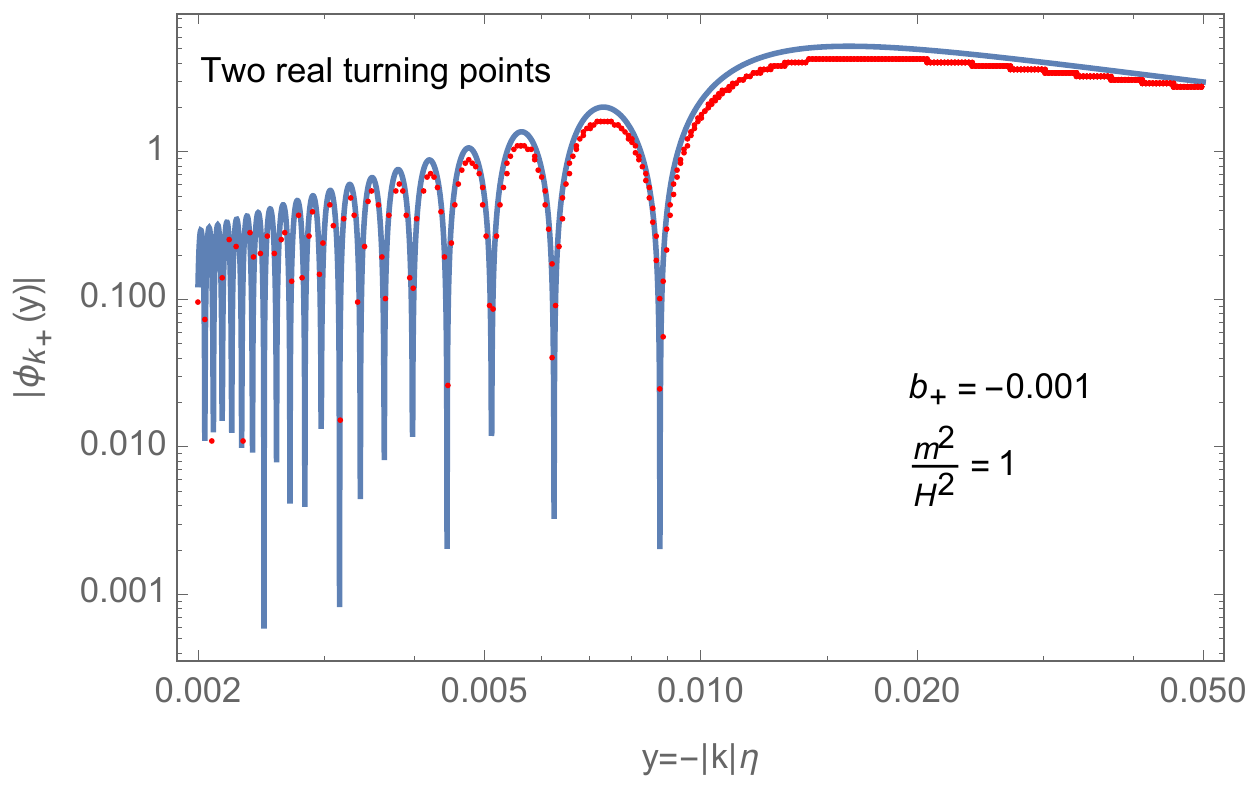}}
{\includegraphics[width=8.0cm]{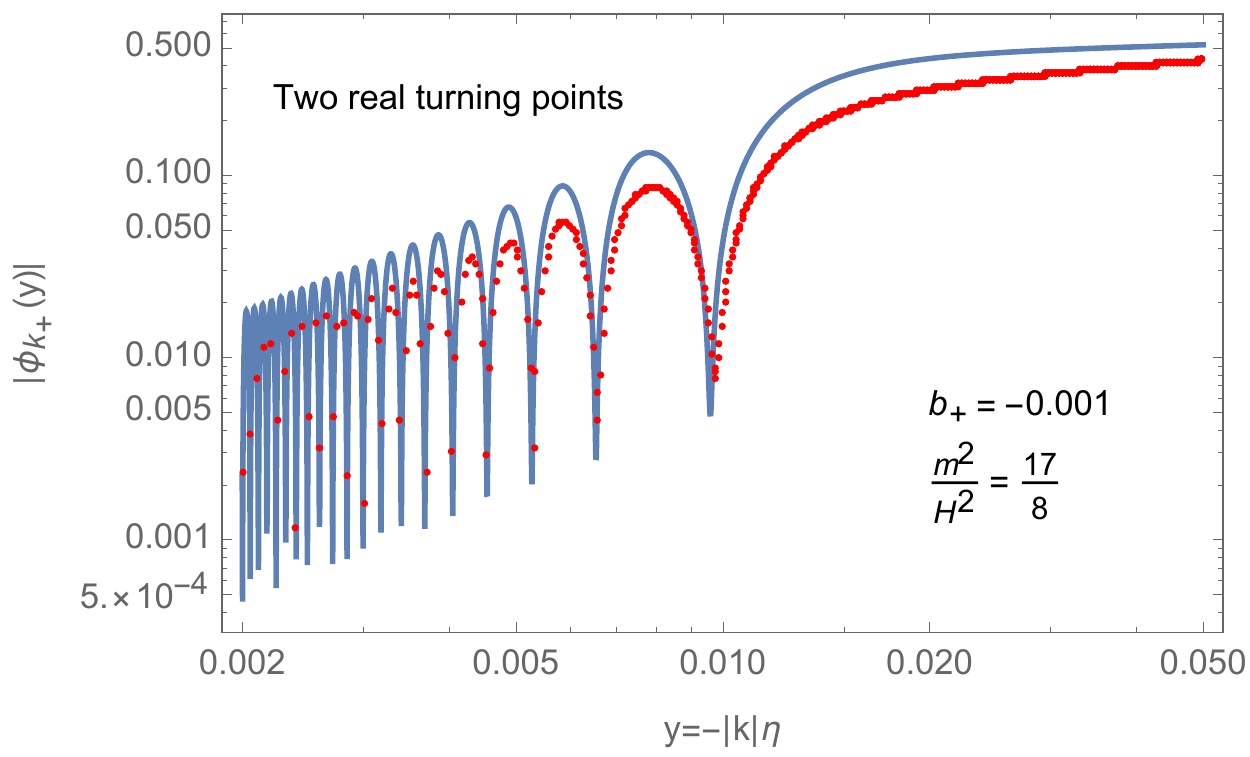}}\\
{\includegraphics[width=8.0cm]{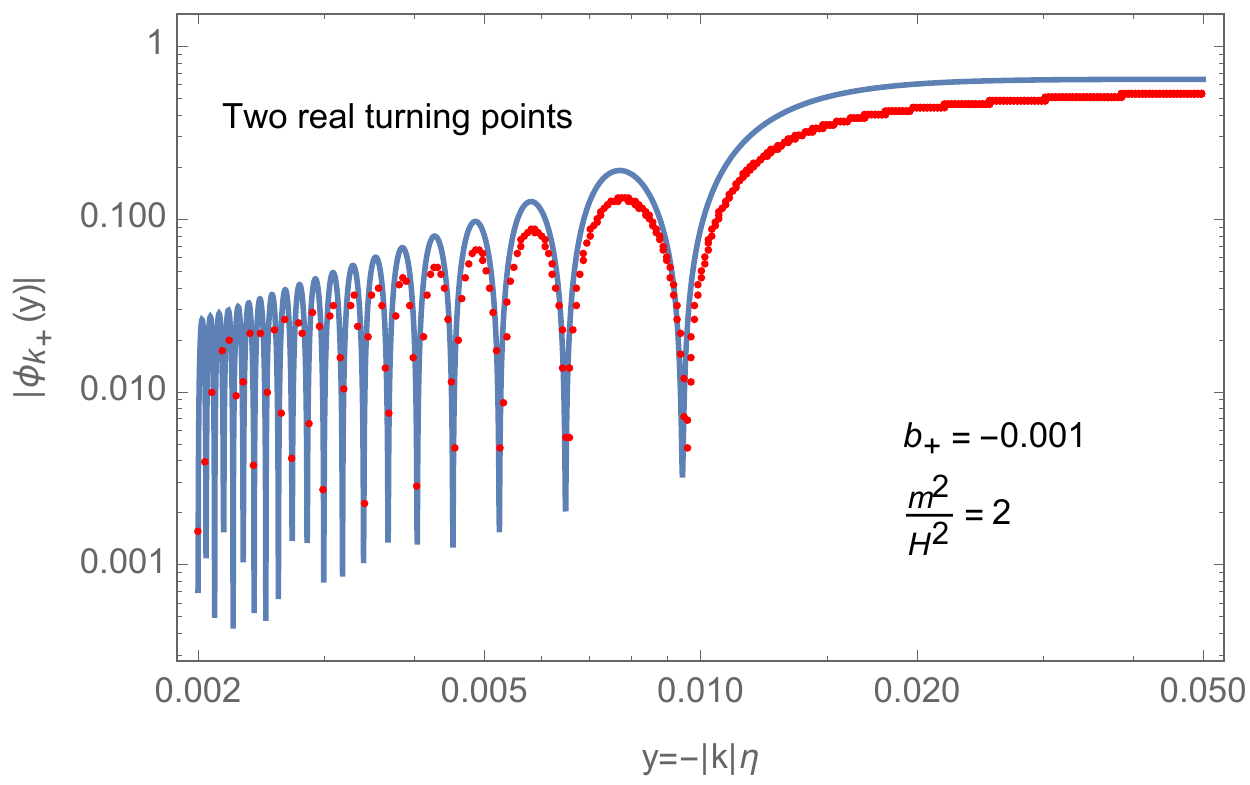}}\\
\caption{Comparison between the analytical (red dotted curve) and numerical solutions (blue solid curve) of the mode function $\phi_{k_+}(y)$($k>0$). The left up panel corresponds to Case (a) with two real and single 
turning points  as shown in Fig.~\ref{gofy_figure_1}, here $\beta=-2 (m^2/H^2=1/4)$, $d=3/2 (p_a^2/k^2=1/2)$, while the right up panel corresponds to Case (c) with two complex turning points as shown in Fig.~\ref{gofy_figure_1}, 
here  $\beta=-3/2\; (m^2/H^2=3/4)$, $d=6(p_a^2/k^2=5)$. The left middle panel corresponds to Case (a)  in Fig.~\ref{gofy_figure_1}, but with a small value of the parameter $b_+=-2k^2E/(3H^4)$, that is, 
$\beta=-5/4 (m^2/H^2=1)$, $b_+=-0.001$, $d=3/2 (p_a^2/k^2=1/2)$. The right middle  panel corresponds to the parameters $b_+=-0.001$, $\beta=-1/8\; (m^2/H^2=17/8)$, and $d=1.5 (p_a^2/k^2=0.5)$.
The bottom panel corresponds to the parameters $b_+=-0.001$, $\beta=-1/4\; (m^2/H^2=2)$,  and $d=1.5 (p_a^2/k^2=0.5)$.} \label{COMPA}
\end{figure}

\begin{figure}
{\includegraphics[width=8.0cm]{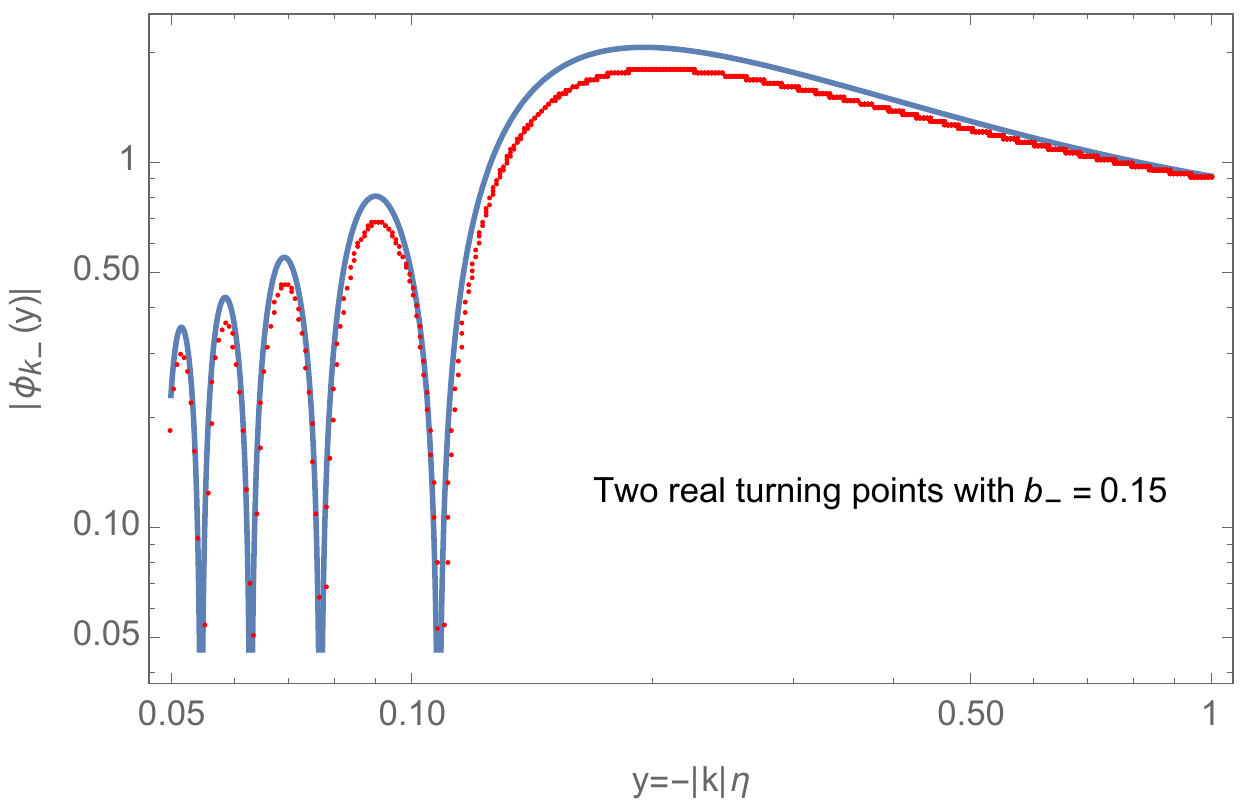}}
{\includegraphics[width=8.0cm]{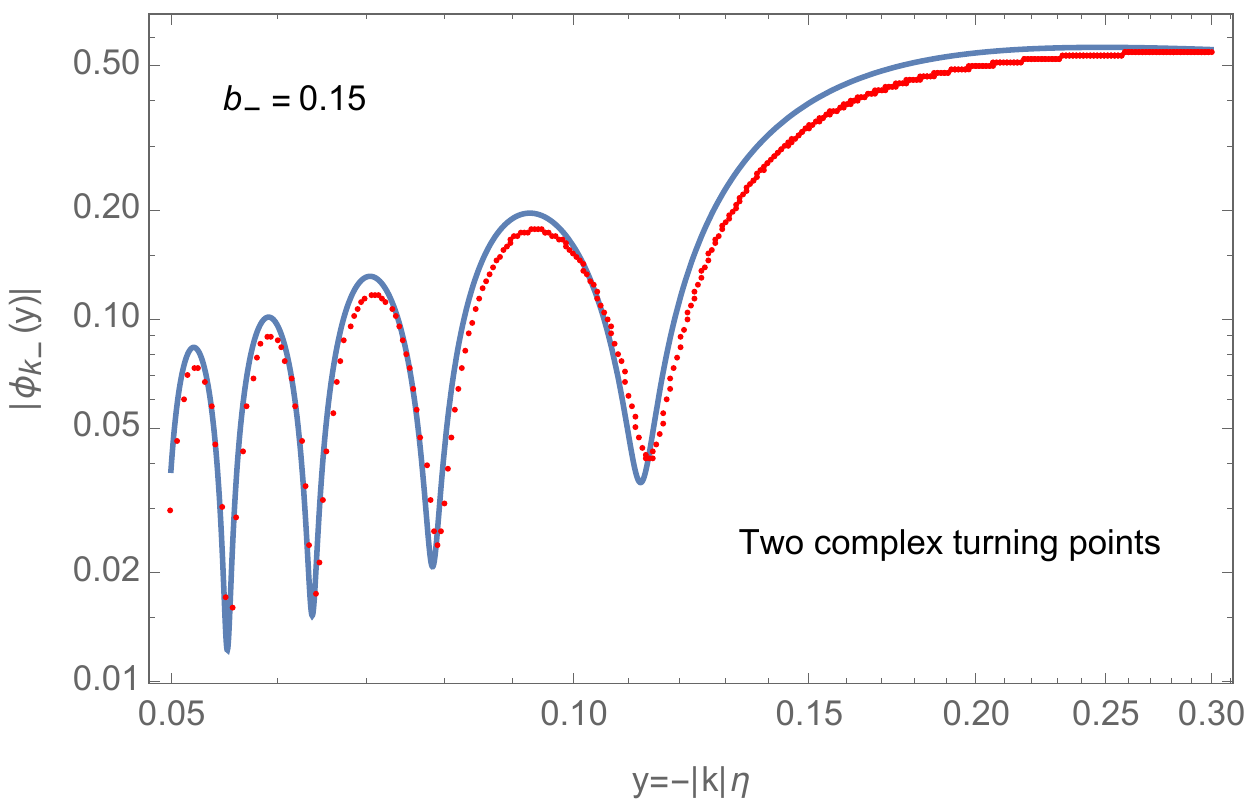}}\\
{\includegraphics[width=8.0cm]{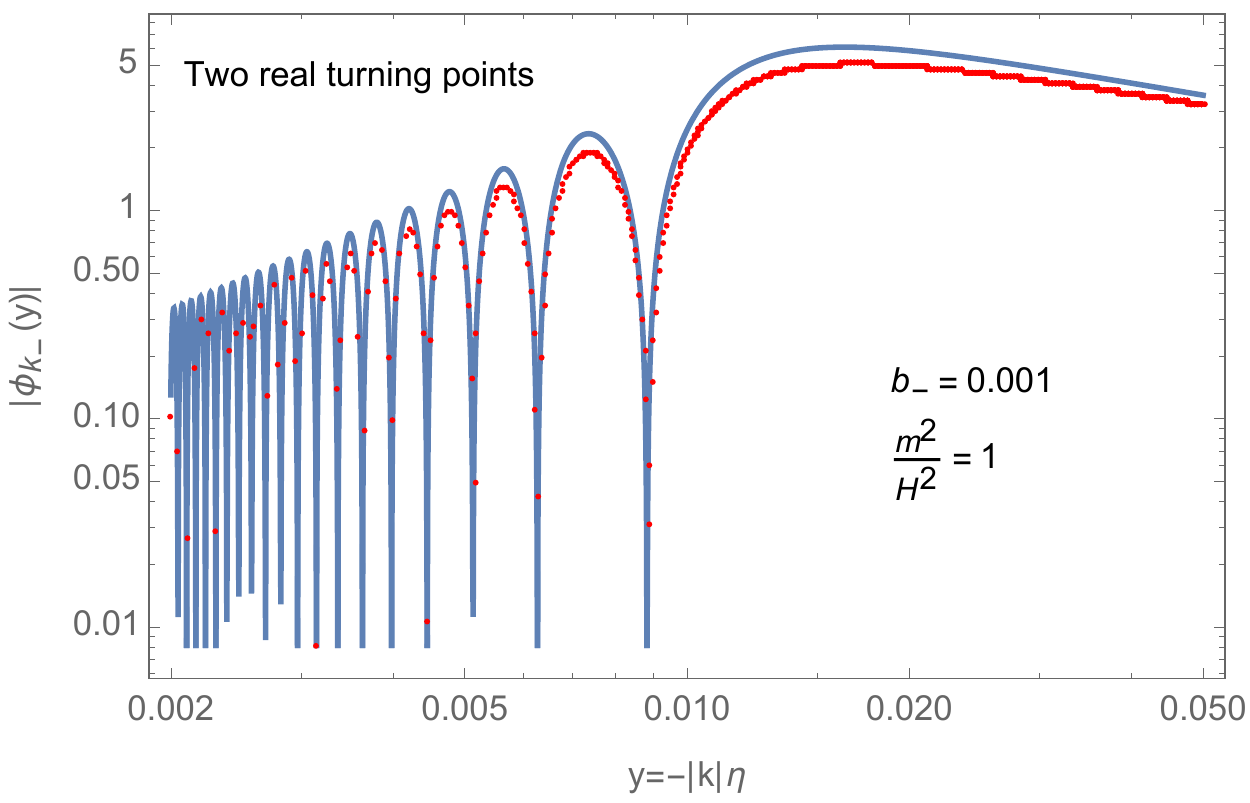}}
{\includegraphics[width=8.0cm]{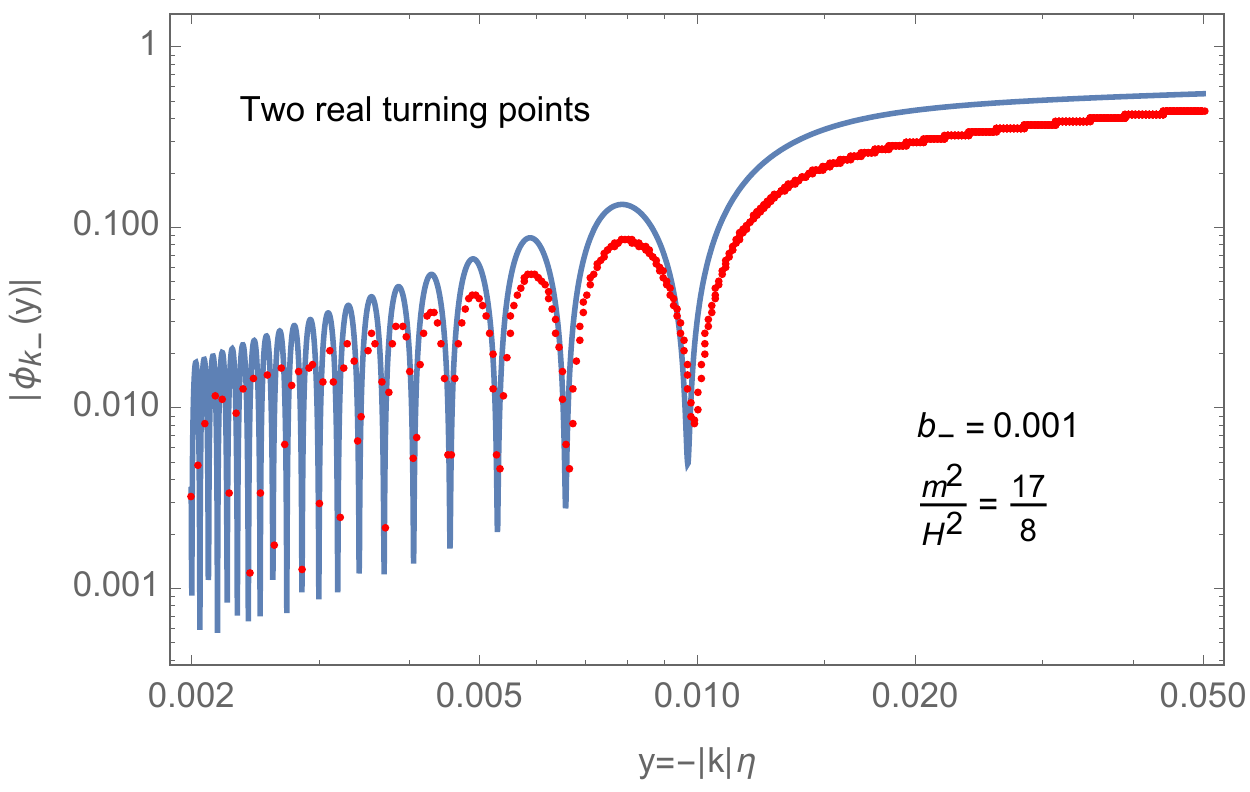}}\\
{\includegraphics[width=8.0cm]{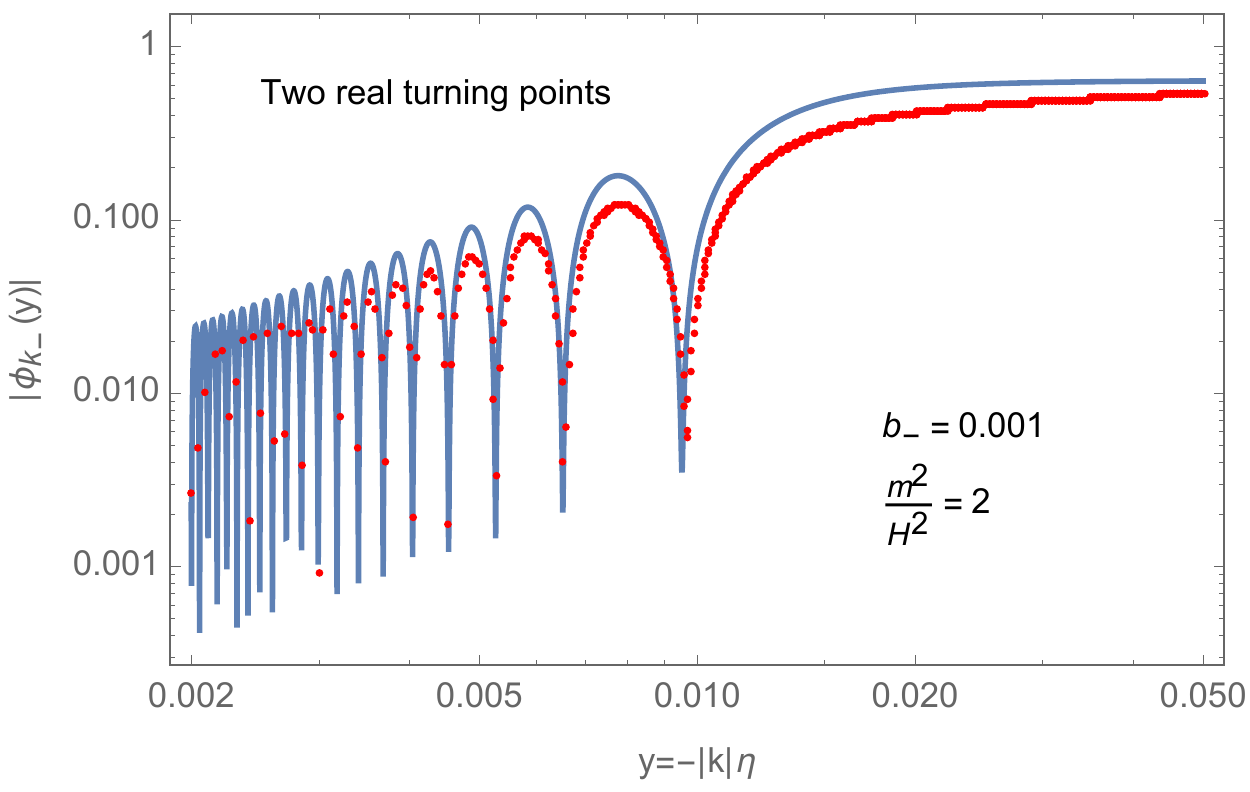}}\\
\caption{Comparison between the analytical (red dotted curve) and numerical solutions (blue solid curve) of the mode function $\phi_{k_-}(y)$($k<0$). The left up panel corresponds to 
Case (a) with two real and single turning points  as shown in Fig.~\ref{gofy_figure_1}, here $\beta=-2 (m^2/H^2=1/4)$, $d=1.05 (p_a^2/k^2=0.05)$, while the right up panel corresponds to 
Case (c) with two complex turning points as shown in Fig.~\ref{gofy_figure_1}, here $\beta=-3/2\; (m^2/H^2=3/4)$, $d=6 \; (p_a^2/k^2=5)$. The left middle  panel corresponds to Case (a) 
with a small value of the parameter $b_-=0.001$ and $\beta=-5/4\; (m^2/H^2=1)$, $d=1.05 (p_a^2/k^2=0.05)$.  The right middle  panel corresponds to 
$b_-=0.001$, $\beta=-1/8\; (m^2/H^2=17/8)$, and $d=1.05 (p_a^2/k^2=0.05)$. The bottom panel corresponds to $b_-=0.001$, $\beta=-1/4\; (m^2/H^2=2)$, and $d=1.5 (p_a^2/k^2=0.5)$.} \label{COMPAB}
\end{figure}

Now let us turn to consider the limit $y\to 0^+$ ($\zeta \to - \infty$). Then,  the asymptotic form of the parabolic cylinder function reads
\bqn
W(\text{\textonehalf}\lambda\zeta_0^2,\sqrt{2\lambda}\zeta)=\left(\frac{ 2 j^{-2}(\sqrt{\lambda}\zeta_0)}{\lambda (\zeta^2-\zeta_0^2)}\right)^{1/4}\sin{\mathfrak D},\;\;\;\; 
W(\text{\textonehalf}\lambda\zeta_0^2,-\sqrt{2\lambda}\zeta)=\left(\frac{ 2 j^{2}(\sqrt{\lambda}\zeta_0)}{\lambda (\zeta^2-\zeta_0^2)}\right)^{1/4}\cos{\mathfrak D}.\nb\\
\eqn
Here
\bqn
\mathfrak{D}=-\lambda \int_{\text{Re}(y_1)}^y \sqrt{-\hat g} dy+\frac{\pi}{4}+\phi\left(\frac{\lambda}{2}\zeta_0^2\right).
\eqn
With these expressions for $y\to 0^+$ we find that the approximate solution can be cast in the form
\bqn\lb{asymptotic}
\phi_{k_\pm}(y)&\simeq& \alpha_1 \left(\frac{2 j^{-2}}{-\lambda \hat g}\right)^{1/4}  \sin{\mathfrak{D}}+\beta_1  \left(\frac{2 j^{2}}{-\lambda \hat g}\right)^{1/4}\cos{\mathfrak{D}}\nb\\
&=& \frac{\alpha_{k_\pm}}{\sqrt{2\lambda }(-\hat g)^{1/4}}e^{-i \mathfrak{D}}+\frac{\beta_{k_\pm}}{\sqrt{2\lambda }(-\hat g)^{1/4}}e^{i\mathfrak{D}},
\eqn
where
\bqn\lb{akbk}
\alpha_{k_\pm}=\frac{1}{2}\left(j+\frac{1}{j}\right),\;\;\beta_{k_\pm}=\frac{i}{2}\left(j-\frac{1}{j}\right) \ .
\eqn
Note that we have ignored the irrelevant phase factor $e^{i\theta}$. The coefficients  $\alpha_{k_\pm}$ and $\beta_{k_\pm}$ depend on the signs of $k$ through Eqs.(\ref{parameters}), (\ref{general_g}), (\ref{zeta0})
and (\ref{jf}). It is easy to show that $|\alpha_{k_\pm}|^2-|\beta_{k_\pm}^2|=1$. In Figs.~\ref{COMPA} ($k > 0$)  and \ref{COMPAB} ($k < 0$) we display both numerical  and analytical solutions  
 with $\alpha_{k_\pm}$ and $\beta_{k_\pm}$ being given by Eq.(\ref{akbk}). From these figures we can see that the analytical solutions trace  the numerical ones very well.

\section{Schwinger Pair Production in de Sitter space}
\renewcommand{\theequation}{4.\arabic{equation}} \setcounter{equation}{0}

Now we are in a position to calculate the Schwinger pair production rate  $|\beta_{k}^2|$ (here $k$ can be either positive or negative).
 Using eqs.(\ref{jform}) and (\ref{akbk}), we find that 
\bqn
|\beta_k|^2 &=& e^{\pi \zeta_0^2} \ .
\eqn
Here $\beta_k$ is the Bogoliubov coefficient that measures the particle production rate. From the above equation, we can see that  
 $|\beta_k^2|$ is determined by the quantity $\zeta_0^2$. Before proceeding further, several remarks about the nature of $\zeta_0^2$ now are in order. 
 First, the sign of $\zeta_0^2$ is sensitive to the nature of the turning points $y_1$ and $y_2$, for which we classify them into several classes.  
\begin{itemize}
\item When $y_1$ and $y_2$ are both single and real, we have $\zeta_0^2>0$, which implies that the particle production during the process is exponentially enhanced. 
As we have shown in Sec.~\ref{turning_points}, for this case to happen, one must  require $\beta = m^2/H^2-9/4<0$.
\item When $y_1$ and $y_2$ are two real but equal roots, i.e., $y_1=y_2$, we have $\zeta_0^2=0$. Then,  we have
\bqn
|\beta_k|^2=1  \ .
\eqn
\item When $y_1$ and $y_2$ are complex conjugate, i.e., $y_1=y_2*$, $\zeta_0^2$ is negative. This implies that the particle production is exponentially suppressed.
\end{itemize} 
In the following we are going to calculate $\zeta_0^2$ explicitly for the several specific cases considered above.

\subsubsection{\bf $\beta=0$} 

\begin{figure}
{\includegraphics[width=8.1cm]{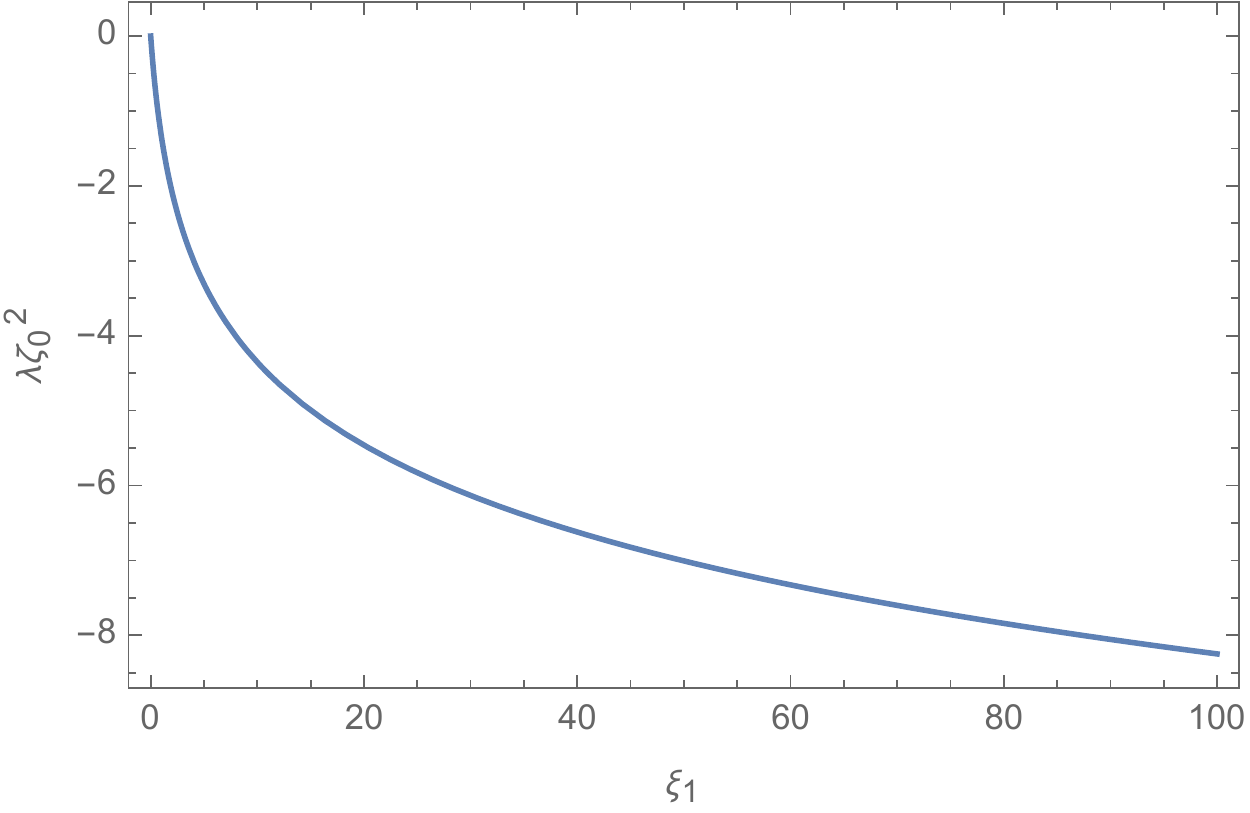}}
\caption{In this figure, $\sqrt{d}\xi_2/6$ is set to one, we can see $\zeta^2_0$ is always negative. In the special case when $p_a/k=1$, the particle creation rate is mode dependent as $|\beta_k|^2=e^{-\alpha k^{2/3}}$ with $\alpha$ being a constant determined by Hubble parameter H and electric field strength E.} \label{aaa}
\end{figure}

In this case, as we showed in Sec.~\ref{turning_points}, the turning points $y_1$ and $y_2$ can be explicitly obtained, which are given by Eq.(\ref{y1y2_caseA}). Then, $\zeta_0^2$ for both positive and negative wavenumber is given by
\bqn
\lambda \zeta_0^2&=&\frac{2}{\pi}\int_{y_1}^{y_2} \sqrt{g}dy=  \frac{2 \sqrt{d}}{\pi} \int_{y_1}^{y_2} \frac{\sqrt{-(y^3-y_1^3)(y^3-y_2^3)}}{y^3}dy\nb\\
&=& \frac{\sqrt{d}\xi_1\xi_2}{6(\xi_1+1)} \Bigg\{
 \Big(3\xi_1+8\Big)\,_2F_1\left(\frac{1}{3},\frac{5}{6};2;-\xi_1\right)-10\, _2F_1\left(-\frac{2}{3},\frac{5}{6};2;-\xi_1\right)\Bigg\}.
\eqn
Here $\xi_1=p^2_a/k^2$ and $\xi_2=(\frac{k^2E}{3H^4d})^{1/3}$ . Fig.~\ref{aaa} shows that $\lambda \zeta_0^2$ is always negative. Thus, we have
\bqn
|\beta_k|^2 = e^{\pi \zeta_0^2} <1.
\eqn
That is, in the current case the particle production is always suppressed. In addition, the above results show that $\lambda \zeta_0^2 \propto - E^{1/3}$, which implies that the particle production rate becomes smaller when the electric field becomes larger. This seems in contradiction with the results obtained in \cite{Frob:2014zka}, where the particle production rate is enhanced when $E$ becomes larger for the $k<0$ modes. Actually,  both results are correct, and  can be understood qualitatively by using the WKB analysis, which is presented in detail in Appendix E, in which we show that for the case considered  in this subsection, the larger electric field tends to make the WKB condition well satisfied, while the case studied  in \cite{Frob:2014zka} tends to make the WKB condition violated or not well satisfied. This explains why the particle production rate is suppressed by large electric field in the current case, while it is enhanced for the case considered in \cite{Frob:2014zka}. For details, we refer readers to Appendix E. 

\subsubsection{Weak Electric Field Limit}

\begin{figure}
{\includegraphics[width=8.1cm]{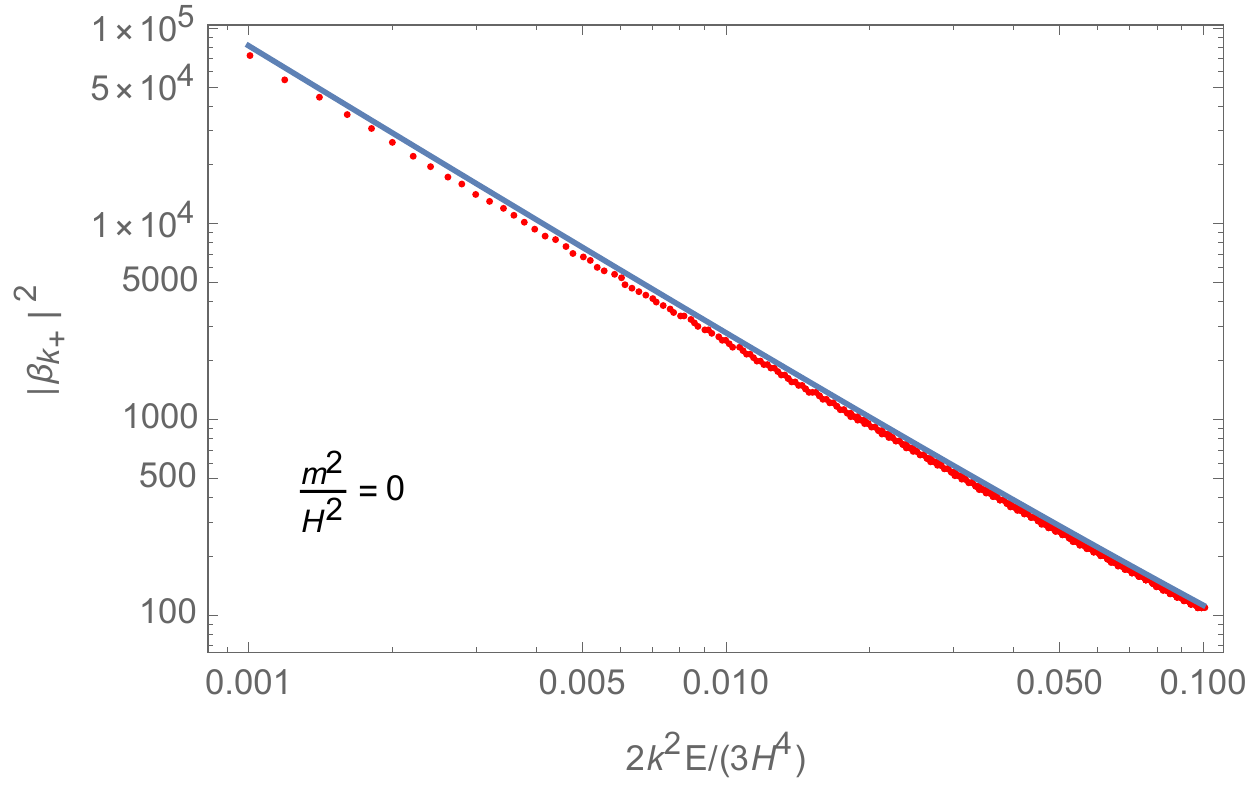}}
{\includegraphics[width=8.1cm]{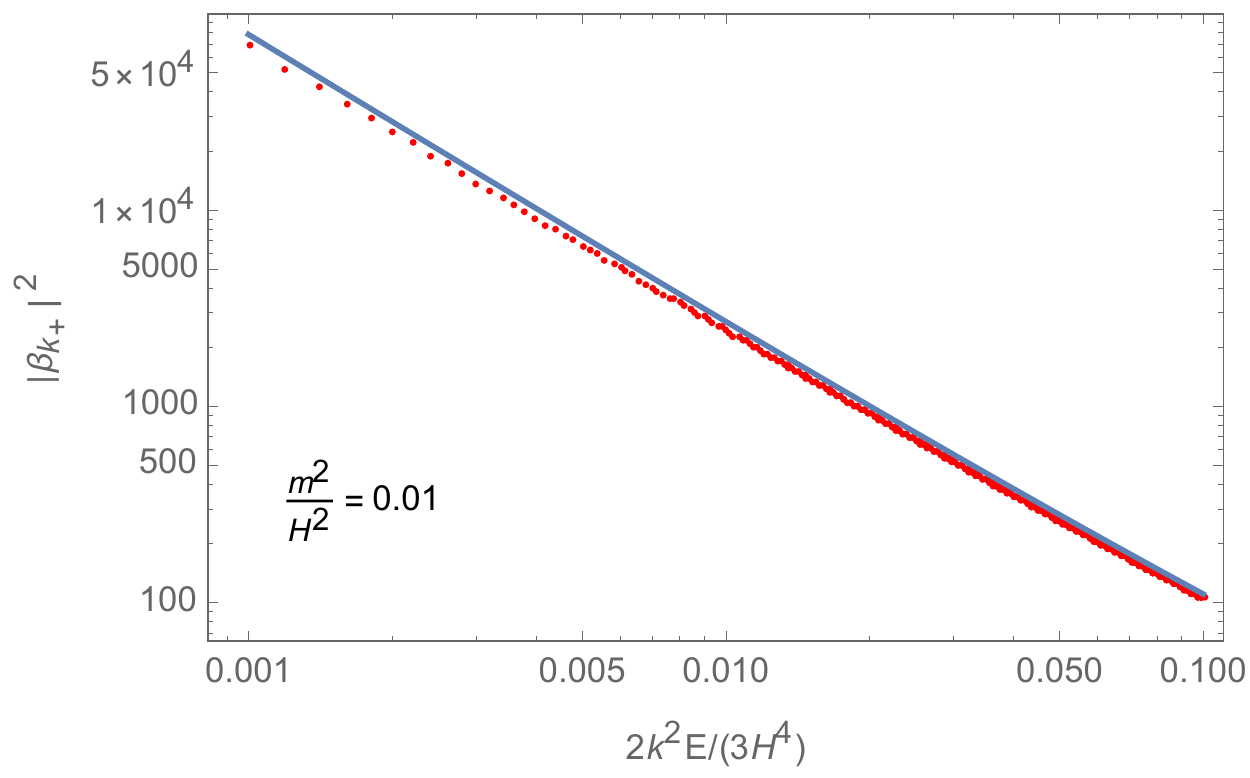}}\\
{\includegraphics[width=8.1cm]{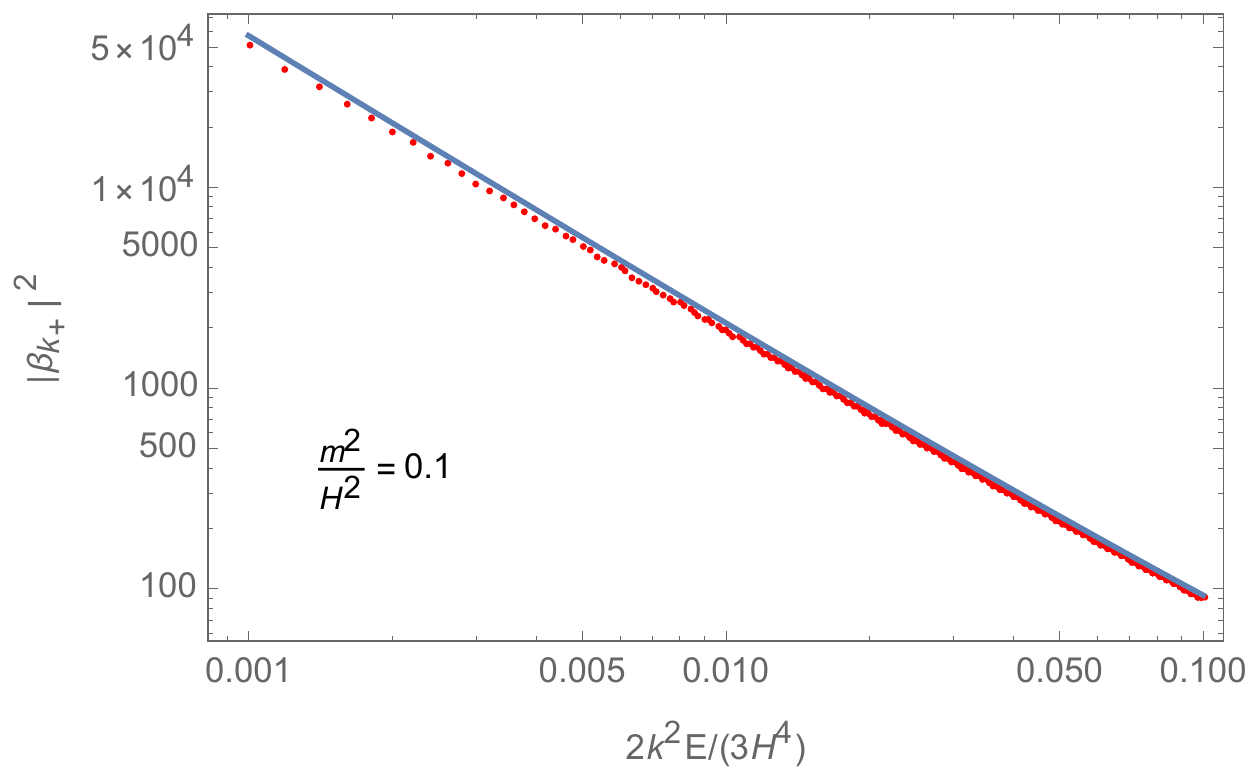}}
{\includegraphics[width=8.1cm]{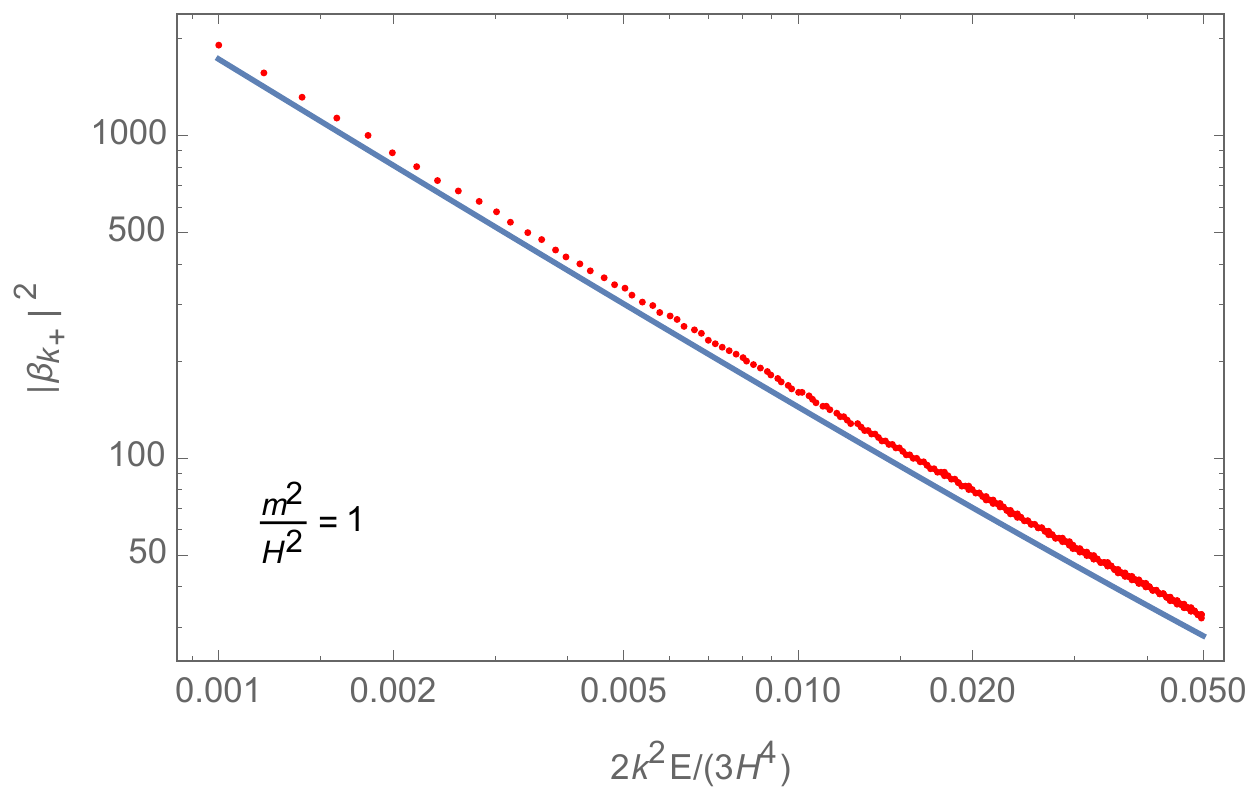}}\\
\caption{These figures shows  the behavior of $|\beta_{k_+}|^2$ both numerically (red dotted curve) and analytically (blue solid curve) for different 
values of $m^2/H^2$ when the electric field is very weak. Here, we choose $p_a^2/k^2=1$.} \label{3d_2}
\end{figure}

\begin{figure}
{\includegraphics[width=8.1cm]{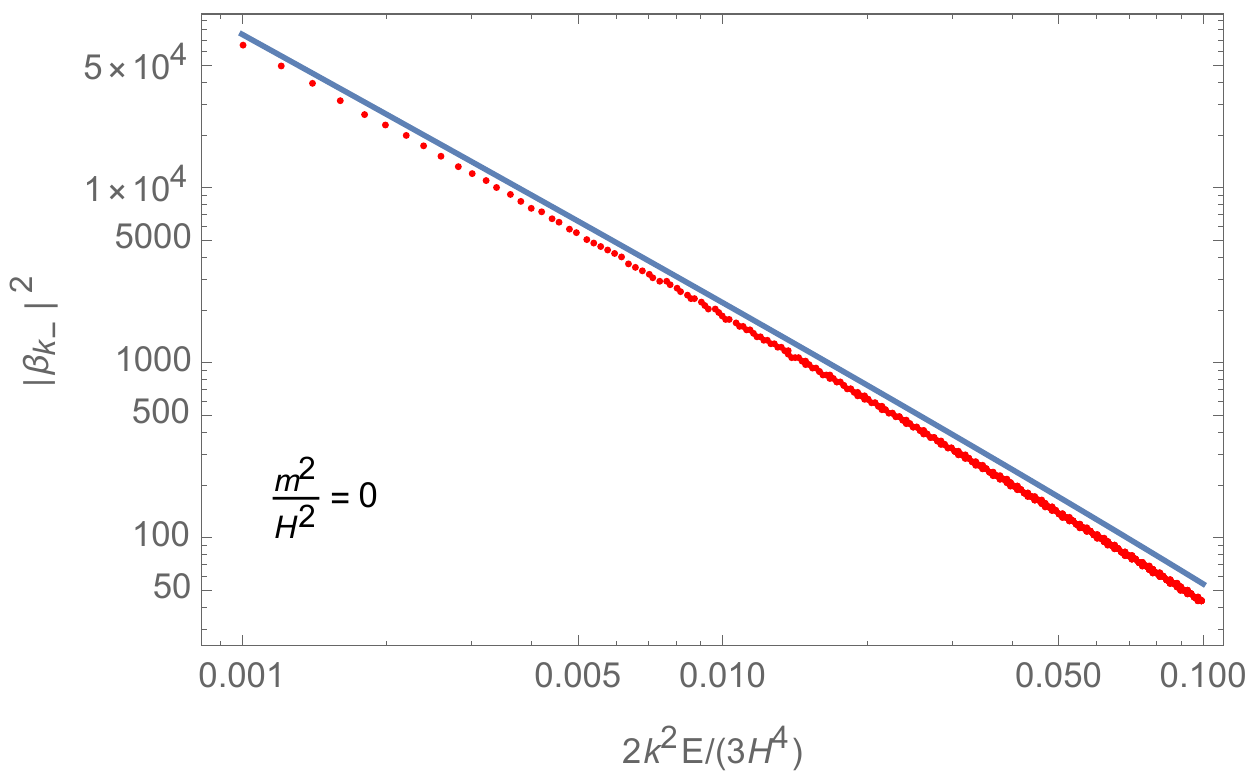}}
{\includegraphics[width=8.1cm]{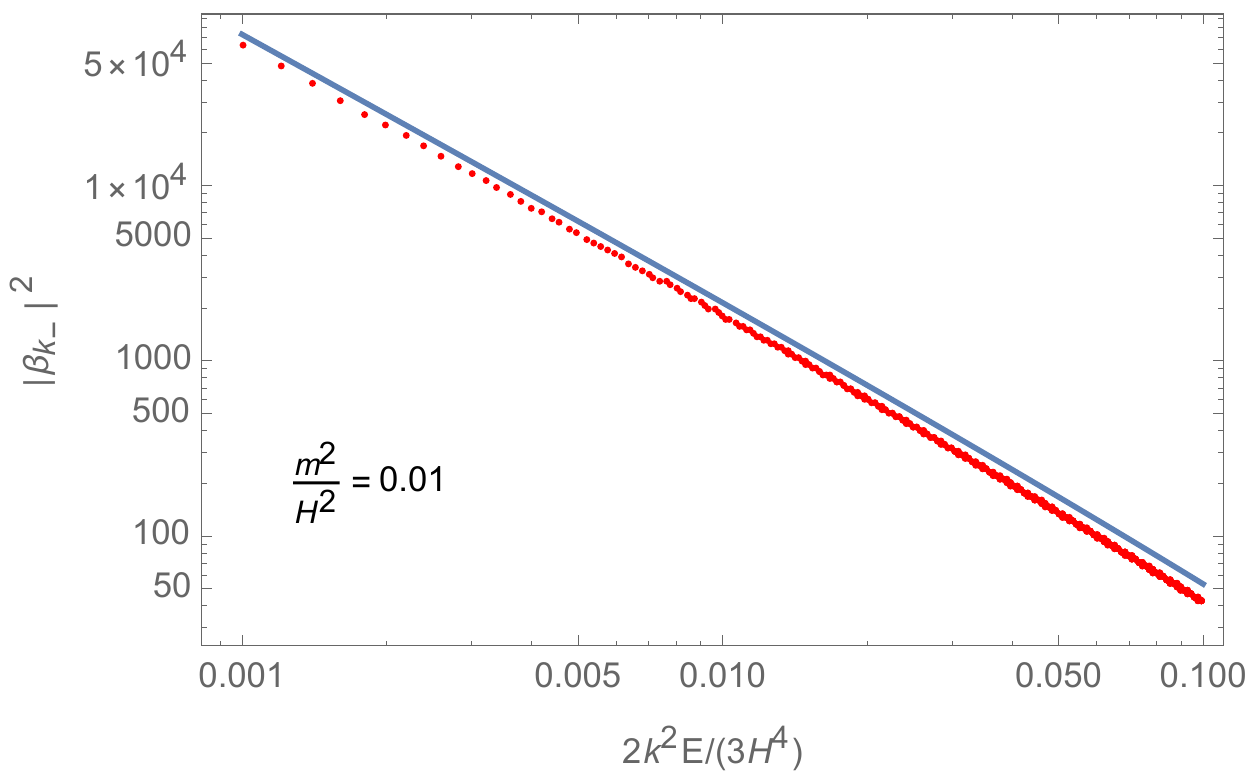}}\\
{\includegraphics[width=8.1cm]{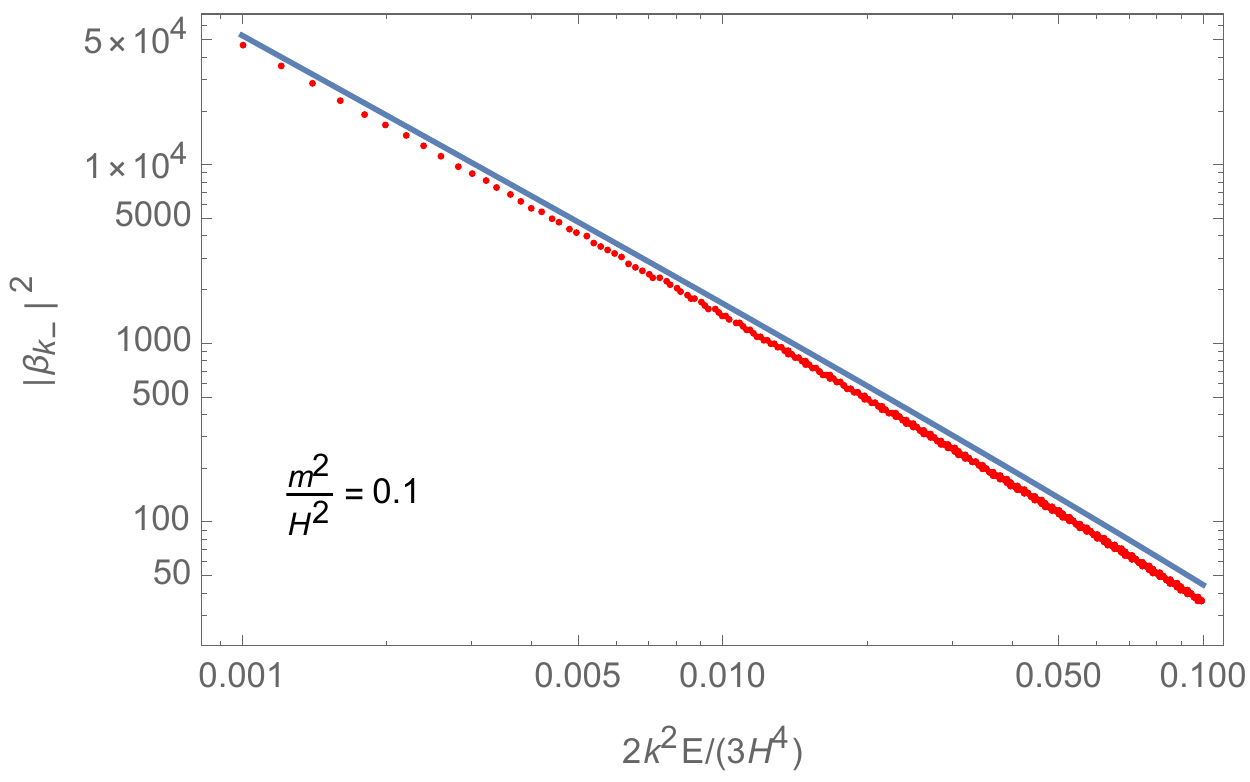}}
{\includegraphics[width=8.1cm]{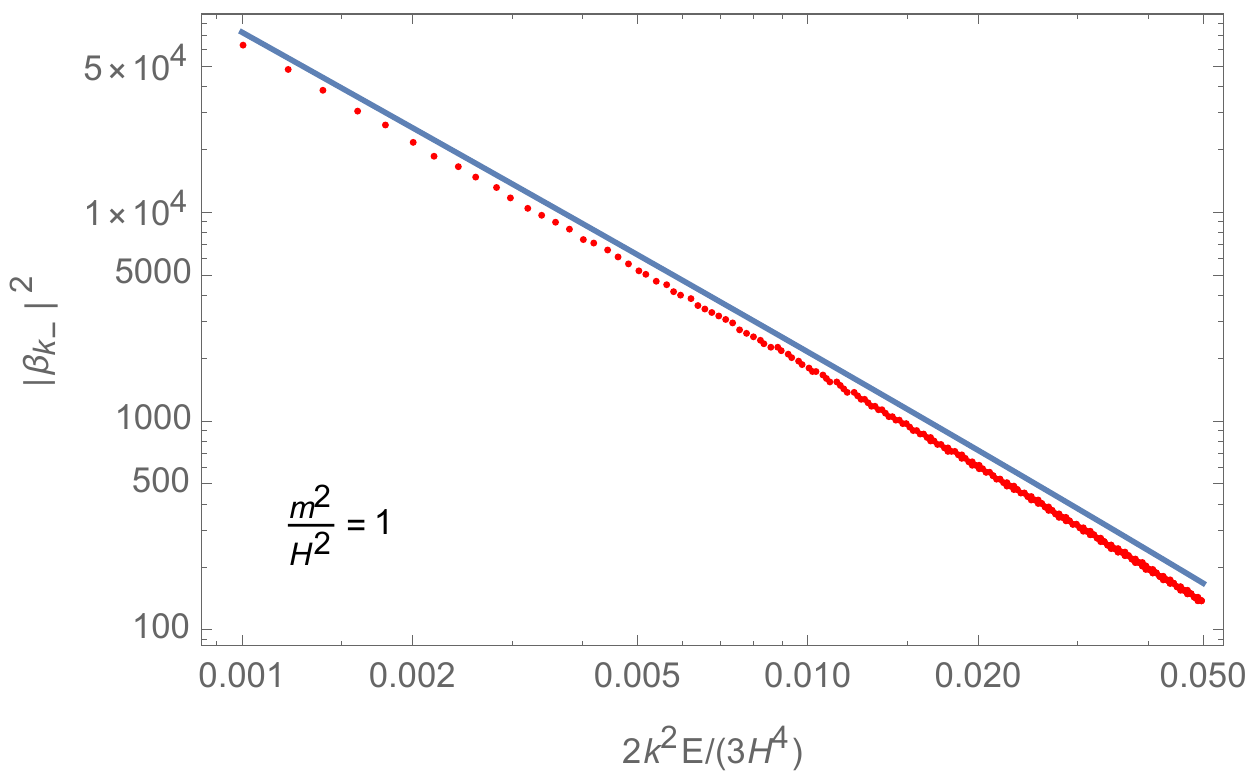}}\\
\caption{These figures shows  the behavior of $|\beta_{k_-}|^2$ both numerically (red dotted curve) and analytically (blue solid curve) for 
different values of $m^2/H^2$ when the electric field is very weak. Here, we choose $p_a^2/k^2=1$.} \label{3d_3}
\end{figure}

When the electric field is weak, that is, when $\frac{2k^2E}{3H^4} \ll 1$, both $b_\pm$ and $c$ are small quantities according to Eq.(\ref{eom}), so we can find approximate solution of $\zeta^2_0$ in the powers of $b_\pm$. 
First, for  two turning points of Eq.(\ref{roots}), $y_{1,2}$  can be derived by expanding them in terms of  $b_\pm$ and are given explicitly by Eqs.(\ref{y1y2_CaseC}) and (\ref{y1y2_CaseD}).
Since 
\bq
 \zeta_0^2=\frac{2}{\pi}\int_{y_1}^{y_2} \sqrt{g}dy,
\eq
we also need to expand the integrand in powers of $b_\pm$, which is 
\bq
g(y)=\frac{\sqrt{-\beta-d y^2}}{y}-\frac{b_\pm}{2y^2\sqrt{-\beta-d y^2}}+\mathcal{O}\Big(|b_\pm|^2\Big).
\eq
Here we assume   $\beta < 0$,  since we already know that  the particle creation rate is suppressed for  $\beta > 0$. 
Then, straightforward calculations yield 
\bqn
\pi \zeta_0^2&=&2 \left[ \frac{\sqrt{-\beta}}{4}\left(-4+\ln \left(\frac{-64\beta^3}{d^2}\right)-2\text{ln}(|b_\pm|)\right)\pm\frac{\sqrt{|b_\pm|}}{\sqrt{2}(-\beta)^{1/4}} +\mathcal{O}\big((|b_\pm|)\big)\right]\nb\\
&=& \sqrt{\frac{9}{4}-\frac{m^2}{H^2}}\left[3\ln 2-2+\frac{3}
{2} \ln \left(\frac{9}{4}-\frac{ m^2}{H^2}\right)- \ln \left(1+\frac{p_a^2}{k^2}\right)-\ln\left(\frac{2 k^2E}{3H^4}\right)\right]\nb\\
&& \pm\frac{2}{\sqrt{3}} \frac{|k| E^{1/2}}{H^2} \left(\frac{9}{4}-\frac{m^2}{H^2}\right)^{-1/4}  +\mathcal{O}\big(E\big).
\eqn
Thus, the Bogoliubov coefficient $|\beta_{k_\pm}|^2$ is given by
\bqn
|\beta_{k_\pm}|^2 = e^{\pi \zeta_0^2(b_\pm)} \propto  \left(\frac{3H^4}{2 k^2E}\right)^{ \sqrt{\frac{9}{4}-\frac{m^2}{H^2}}},
\eqn
which diverge as $k^2E/H^4 \rightarrow 0$, and are plotted out for various values of $m^2/H^2$ in Figs. \ref{3d_2} and \ref{3d_3}.
  In these  figures, we also plot  the numerical  solutions, which show that the
approximate analytical solutions trace the numerical ones very well (see Appendix~\ref{numerical} for how to compute the numerical solutions). 
 Then,  it is easy to see that in the weak electric field limit, $E\approx 0$, the electric current $J$ is given by  \cite{Frob:2014zka}
\bqn
\lb{currentJ}
J \propto\Big( |\beta_{k_+}|^2-|\beta_{k_-}|^2 \Big)\approx {\cal{A}}  \left(\frac{k^2E}{H^4}\right)^{\frac{1}{2}- \sqrt{\frac{9}{4}-\frac{m^2}{H^2}}}  +\mathcal{O}\left(E^{\delta}\right),
\eqn
where   
\bq
{\cal{A}} \equiv -\frac{4}{\sqrt{3}}(|\beta|)^{-1/4}e^{-2\sqrt{|\beta|}}\left(\frac{12|\beta|^{3/2}}{d}\right)^{\sqrt{|\beta|}}, \;\;\;\;
\delta \equiv \frac{3}{2}- \sqrt{\frac{9}{4}-\frac{m^2}{H^2}} \ge 0.
\eq
Note that the above current is the semiclassical one.
The derivation is almost pararell to that in \cite{Frob:2014zka}. 
Unfortunately, we could not compare this semi-classical current with a renormalized one,  because we do not know the analytic expression for the mode functions.
However, in the current situation, we believe the semi-classical analysis is sufficient to get a qualitative understanding.
Recall that  $\beta=m^2/H^2-9/4 < 0$ and $d=p^2_a / k^2+1$.
From the above expressions, we find that there exists a critical mass $m_c$,
\bq
\lb{cmass}
m_{c}\equiv \sqrt{2}H,
\eq
and for $m < m_c$,  the electric current  $J$ diverges   as ${k^2E}/{H^4} \rightarrow 0$. This is similar to what was found in \cite{Frob:2014zka},  the so-called  {\em infrared hyperconductivity}.
However,  in the (1+1)-dimensional case  this occurs only when $m = 0$, and the corresponding  electric current $J$  is inversely proportional to $E$  \cite{Frob:2014zka}. 
It is interesting to note that $J \propto E^{-1}$ for $m = 0$ even in the (1+3)-dimensional de Sitter background,  as can be seen from Eq.(\ref{currentJ}). But, for $m \in (0, m_c)$, it diverges as
$1/E^{\Delta}$,  where $\Delta \equiv \sqrt{\frac{9}{4}-\frac{m^2}{H^2}}- \frac{1}{2}$ and $\Delta \in (0, 1)$.

\section{Conclusion}

In this paper, we studied the Schwinger pair production process due to the persistent electric field in the (3+1)-dimensional  de Sitter space.
The persistent electric field can be sustained by the inflaton during anisotropic inflation~\cite{Watanabe:2009ct}.
More precisely, we analytically investigated  the Schwinger effect in  the de Sitter space by using the uniform asymptotic approximation method, 
and showed that the equation of motion in general has two turning points.
 The nature of these points could be single, double, real or complex, depending on the choice of the free parameters involved in the theory.  
Those different natures lead to different rates of particle production. For example, it is enhanced 
when both turning points are real and single, and suppressed when both are complex. Furthermore, for the turning points to be real, it is necessary 
$\beta \equiv m^2/H^2-9/4 < 0$, and  the more negative,  the easier to produce particles. 
When $\beta$ is positive, both turning points are always complex,  
and the particle production due to the Schwinger process is exponentially suppressed  for massive modes like as the conventional Schwinger effect. 
 
We focussed on the case $\beta < 0$ and studied the 
particle pair production for $b \; [\equiv 2k^2E/(3H^4)]$  very small. 
Remarkably, we found  that the particle production rate is  $|\beta_k^2| \propto 1/E^{\sqrt{|\beta|}}$, 
which diverges  as $E \to 0$, that is, the particle production is strongly enhanced in the weak electric field limit. 
This indicates the occurrence of hyperconductivity in anisotropic inflation. This behavior originates from
the infrared behavior of light fields in de Sitter space. In fact, it is well known the massless field shows the bad behavior
in the infrared. In the presence of the electric field, the divergence is regulated. However, as the electric field decreases,
the infrared divergence shows up in the form of the hyperconductivity. Thus, for a scalar field with the finite mass, 
this apparent divergence should be regulated at some point.  It is interesting to note that   the semi-classical current is given by Eq.(\ref{currentJ}), 
which also diverges when the mass $m$ of the emitted particles is less than the critical mass
$m_c$ defined by Eq.(\ref{cmass}). It would be very interesting to compare the semi-classical current
 with the renormalized one. We leave this for future work.

\acknowledgments
The authors would like to thank Gerald Cleaver, Klaus Kirsten and Qin Sheng for valuable comments and suggestions. 
Part of the work was done when J.J.G was visiting Physics Department, Baylor University (BU). She would like to express her gratitude to the Department and BU for their hospitality.
The work was  supported in part by JSPS KAKENHI Grant Number 17H02894 and MEXT KAKENHI Grant Number 15H05895 (J.S.), Baylor graduate programs 
through Physics Department (B.F.L.), National Natural Science Foundation of China (NNSFC),  Grants Nos. 11375153 (A.W.),
11675145(A.W.), 11205133 (Q.W.), 11675143 (J.J.G, T.Z.), 11205133 (Q.W., T.Z.) and 11105120 (T.Z.), and the 2016 Postdoctoral  Fellowship of Zhejiang Province, China (J.J.G.). 


\appendix

\section{Uniform Asymptotic Approximation with two turning points}
\renewcommand{\theequation}{A.\arabic{equation}} \setcounter{equation}{0}
\lb{Uniform_App}

Now let us apply the uniform asymptotic approximation method  to obtain approximate analytical  solutions of $\phi_k(y)$. To proceed, let us introduce the Liouville transformation with two new variables $U(\zeta)$ and $\zeta$,
%
\bqn\lb{Olver trans}
U(\zeta)=\chi^{1/4} \phi_k(y), \;\;\;\chi=\left(\frac{d\zeta}{dy}\right)^2=\frac{|\hat{g}(y)|}{f^{(1)}(\zeta)^2},
\eqn
where
\bqn
f(\zeta)=\int \sqrt{|\hat{g}(y)|}dy,\;\;\;f^{(1)}(\zeta)=\frac{df(\zeta)}{d\zeta}.
\eqn
Note that $\chi$ must be regular and not vanish in the intervals of interest. Consequently, {\em the function $f(\zeta)$ must be chosen so that $f^{(1)}(\zeta)$ has zeros and singularities of the same type as those of $\hat g(y)$}. As shown below, such requirement plays an essential role in determining the approximate solution of $\phi_k(y)$. In terms of $U(\zeta)$ and $\zeta$, Eq.(\ref{eom_uniform}) changes into the form,
\bqn\lb{eomU}
\frac{d^2 U(\zeta)}{d\zeta^2} =\Big[\pm \lambda^2 f^{(1)}(\zeta)^2+\psi(\zeta)\Big]U(\zeta),
\eqn
where
\bqn
\psi(\zeta)&=&\frac{q(y)}{\chi}-\chi^{-3/4} \frac{d^2 (\chi^{-1/4})}{dy^2} =\frac{q}{\chi}-\frac{5}{16}\frac{1}{\chi^3}\left(\frac{d\chi}{dy}\right)^2+\frac{1}{4}\frac{1}{\chi^2}\frac{d^2\chi}{dy^2},
\eqn
here $\pm$ correspond to $\hat{g}(y)>0$ and $\hat{g}(y)<0$, respectively. Considering $\psi(\zeta)=0$ as the first-order approximation, one can choose $f^{(1)}(\zeta)^2$ so that: (a) the first-order approximation can be as close to the exact solution as possible, and (b) the resulting equation can be solved explicitly (in terms of known functions). Clearly, such a choice sensitively depends on the behavior of the functions $\lambda^2 \hat g(y)$ and $q(y)$ near their poles and zeros.

Following \cite{zhu_high-order_2016}, for a pair of turning points $y_{1}$ and $y_2$, the crucial point is to choose $f^{(1)}(\zeta)^2$ in the Liouville transformations (\ref{Olver trans}) as \footnote{Here we use $\zeta$ to denote the variable $\xi$ in the Liouville transformation, for the purpose to distinguish with the variable $\xi$ used for the $y_0$ case.}
\bqn
\lb{xi0A}
f^{(1)}(\zeta)^2=|\zeta^2-\zeta_0^2|,
\eqn
where we choose $\zeta$ is an increasing function of $y$, and with the choices $\zeta(y_1)=-\zeta_0$ and $\zeta(y_2)=\zeta_0$. When $\zeta$ is real, $\zeta_0=0$, and $\zeta$ is pure imaginary, it corresponds to the cases $y_{1,2}$ are real, $y_1=y_2$, and $y_{1,2}$ are complex conjugated, respectively. It can be proved that $\zeta_0^2$ is calculated as
\bqn
\zeta_0^2=\frac{2}{\pi}\int_{y_1}^{y_2} \sqrt{\hat g(y)}dy,
\eqn
which can be positive, or zero, or even negative.

Now let us turn to derive the relation between $\zeta(y)$ and $y$. Let us first consider the case when $y_1$ and $y_2$ are real. When $y>y_2$, we have $\zeta(y)>\zeta_0$, then from Eq. (\ref{Olver trans}), we find
\bqn
\int_{\zeta_0}^{\zeta}\sqrt{v^2-\zeta_0^2}dv=\int_{y_2}^y\sqrt{-\hat g(y')}dy',
\eqn
which yields
\bqn
&&\int_{y_2}^y\sqrt{-\hat g(y')}dy'
=\frac{1}{2}\zeta\sqrt{\zeta^2-\zeta_0^2}-\frac{\zeta_0^2}{2}\operatorname{arcosh}{\left(\frac{\zeta}{\zeta_0}\right)}.
\eqn
When $y<y_1$, we have $\zeta(y)<-\zeta_0$. Then from Eq. (\ref{Olver trans}) we find
\bqn
\int_{-\zeta_0}^{\zeta}\sqrt{v^2-\zeta_0^2}dv=\int_{y_1}^y\sqrt{-\hat g(y')}dy',
\eqn
which yields
\bqn
&&\int_{y_1}^y\sqrt{-\hat g(y')}dy'
=\frac{1}{2}\zeta\sqrt{\zeta^2-\zeta_0^2}+\frac{\zeta_0^2}{2}\operatorname{arcosh}{\left(-\frac{\zeta}{\zeta_0}\right)}.
\eqn
When $y_1\leq y\leq y_2$, we have $-\zeta_0 <\zeta(y)<\zeta_0$, we find
\bqn
\int_{-\zeta_0}^\zeta \sqrt{\zeta_0^2-v^2}dv=\int_{y_1}^y \sqrt{\hat g(y')}dy',
\eqn
which yields
\bqn
\int_{y_1}^y \sqrt{\hat g(y')}dy'&=&\frac{1}{2}\zeta\sqrt{\zeta^2-\zeta_0^2}+\frac{\zeta_0^2}{2}\arccos\left(-\frac{\zeta}{\zeta_0}\right).\nb\\
\eqn
Now let us turn to consider when $y_1$ and $y_2$ are complex. For this case $\zeta_0^2$ is always negative, thus from Eq. (\ref{Olver trans}) one finds
\bqn
\int_0^\zeta \sqrt{\zeta^2-\zeta_0^2}d\zeta=\int_{\text{Re}(y_1)}^y \sqrt{-\hat g(y')}dy',
\eqn
which yields
\bqn
&&\int_{\text{Re}(y_1)}^y \sqrt{-\hat g(y')}dy'=\frac{1}{2}\zeta\sqrt{\zeta^2-\zeta_0^2}-\frac{\zeta_0^2}{2}\ln\left(\frac{\zeta+\sqrt{\zeta^2-\zeta_0^2}}{|\zeta_0|}\right).
\eqn

Then with $f^{(1)}(\zeta)^2$ given in Eq. (\ref{xi0A}), Eq. (\ref{eomU}) reduces to
\bqn
\lb{eomy1y2}
\frac{d^2U}{d\zeta^2}=\left[\lambda^2 \left(\zeta_0^2-\zeta^2\right)+\psi(\zeta)\right]U.
\eqn
Neglecting the $\psi(\zeta)$ term, we find that  the approximate solutions can be expressed in terms of the parabolic cylinder functions $W(\frac{1}{2}\lambda \zeta_0^2,\pm \sqrt{2\lambda} \zeta)$,
and are given by
\bqn
U(\zeta)&=& \alpha_1 \Bigg\{W\left(\frac{1}{2}\lambda \zeta_0^2, \sqrt{2\lambda}\zeta \right)+\epsilon_5\Bigg\}+\beta_1 \Bigg\{W\left(\frac{1}{2}\lambda \zeta_0^2, -\sqrt{2\lambda }\zeta \right)+\epsilon_6\Bigg\},
\eqn
from which we have
\bqn\lb{solutionW}
\phi_k(y)&=&\alpha_1 \left(\frac{\zeta^2-\zeta_0^2}{-\hat g(y)}\right)^{\frac{1}{4}} \left[W\left(\frac{1}{2}\lambda \zeta_0^2, \sqrt{2\lambda}\zeta \right)+\epsilon_5\right]+\beta_1 \left(\frac{\zeta^2-\zeta_0^2}{-\hat g(y)}\right)^{\frac{1}{4}} \left[W\left(\frac{1}{2}\lambda \zeta_0^2, -\sqrt{2\lambda }\zeta \right)+\epsilon_6\right],\nb\\
\eqn
where $\epsilon_5$ and $\epsilon_6$ are  the errors of the corresponding approximate solutions, which are given by
\bqn
&&\frac{|\epsilon_5|}{M\left(\frac{1}{2}\lambda \zeta_0^2,\sqrt{2\lambda }\zeta\right)},\;\frac{|\partial \epsilon_5/\partial \zeta|}{\sqrt{2} N\left(\frac{1}{2}\lambda \zeta_0^2,\sqrt{2\lambda }\zeta\right)}\leq \frac{\kappa}{\lambda_0 E\left(\frac{1}{2}\lambda \zeta_0^2,\sqrt{2\lambda }\zeta\right)} \Bigg\{\exp{\Big(\lambda \mathscr{V}_{\zeta,a_5}(\mathscr{I})\Big)}-1\Bigg\}. ~~~~~~\nb\\
&&\frac{|\epsilon_6|}{M\left(\frac{1}{2}\lambda \zeta_0^2,\sqrt{2\lambda }\zeta\right)},\;\frac{|\partial \epsilon_6/\partial \zeta|}{\sqrt{2} N\left(\frac{1}{2}\lambda \zeta_0^2,\sqrt{2\lambda }\zeta\right)}\leq \frac{\kappa E\left(\frac{1}{2}\lambda \zeta_0^2,\sqrt{2\lambda }\zeta\right)}{\lambda } \Bigg\{\exp{\Big(\lambda_0 \mathscr{V}_{0,\zeta}(\mathscr{I})\Big)}-1\Bigg\}, ~~~~
\eqn
where $M\left(\frac{1}{2}\lambda \zeta_0^2,\sqrt{2\lambda }\zeta\right)$, $N\left(\frac{1}{2}\lambda \zeta_0^2,\sqrt{2\lambda }\zeta\right)$, and $E\left(\frac{1}{2}\lambda \zeta_0^2,\sqrt{2\lambda }\zeta\right)$ are auxiliary functions of the parabolic cylinder functions, and
\bqn\lb{error_twopoint}
\mathscr{I}(\zeta) \equiv \int \frac{\psi(\zeta)}{\sqrt{|\zeta^2-\zeta_0^2|}}dv
\eqn
is the associated error control function for the approximate solutions near $y_1$ and $y_2$. When $\zeta^2>\zeta_0$ (i.e., $y<y_1$ or $y>y_2$), one finds
\bqn\lb{errorI1}
\mathscr{I}(\zeta)&=&-\int_{\pm \zeta_0}^{\zeta}\left\{\frac{q}{\hat g}-\frac{5}{16}\frac{\hat g'^2}{g^3}+\frac{1}{4}\frac{g''}{g^2}\right\}\sqrt{v^2-\zeta_0^2}dv+\int_{\pm\zeta_0}^{\zeta} \left\{\frac{5\zeta_0^2}{4(v^2-\zeta_0^2)^3}+\frac{3}{4(v^2-\zeta_0^2)^2}\right\}\sqrt{v^2-\zeta_0^2}dv\nb\\
&=&-\int_{y_{2,1}}^{y}\left\{\frac{q}{\hat g}-\frac{5}{16}\frac{\hat g'^2}{g^3}+\frac{1}{4}\frac{g''}{g^2}\right\}\sqrt{-\hat g} dy'+\int_{\pm\zeta_0}^{\zeta} \left\{\frac{5\zeta_0^2}{4(v^2-\zeta_0^2)^{5/2}}+\frac{3}{4(v^2-\zeta_0^2)^{3/2}}\right\}dv,\nb\\
\eqn
where $\pm$ (i.e., or $y_{2,1}$) correspond to the cases when $y<y_1$ and $y>y_2$, respectively. Note that Eq. (\ref{errorI1}) is also valid when $y_1$ and $y_2$ are both complex. When $-\zeta_0<\zeta<\zeta_0$ (i.e., $y_1<y<y_2$), we have
\bqn
\mathscr{I}(\zeta)&=&\int_{ \zeta_0}^{\zeta}\left\{\frac{q}{\hat g}-\frac{5}{16}\frac{\hat g'^2}{g^3}+\frac{1}{4}\frac{g''}{g^2}\right\}\sqrt{\zeta_0^2-v^2}dv-\int_{\zeta_0}^{\zeta} \left\{\frac{5\zeta_0^2}{4(\zeta^2-\zeta_0^2)^3}+\frac{3}{4(\zeta^2-\zeta_0^2)^2}\right\}\sqrt{\zeta_0^2-v^2}dv\nb\\
&=&\int_{y_{2}}^{y}\left\{\frac{q}{\hat g}-\frac{5}{16}\frac{\hat g'^2}{g^3}+\frac{1}{4}\frac{g''}{g^2}\right\}\sqrt{-\hat g} dy'+\int_{\zeta_0}^{\zeta} \left\{\frac{5\zeta_0^2}{4(\zeta_0^2-v^2)^{5/2}}-\frac{3}{4(\zeta_0^2-v^2)^2}\right\}dv.\nb\\
\eqn

\section{Asymptotical expansions of parabolic cylinder functions}
\renewcommand{\theequation}{B.\arabic{equation}} \setcounter{equation}{0}
\lb{Asy_W}

The asymptotic expansion of parabolic cylinder functions $W(\text{\textonehalf}\lambda\zeta_0^2,\pm \sqrt{2\lambda}\zeta)$ and $W(\text{\textonehalf}\lambda\zeta_0^2,\pm \sqrt{2\lambda}\zeta)$ for $\zeta^2-\zeta_0^2 \gg 1$ can be written as \cite{GST}
\bqn
W(\text{\textonehalf}\lambda\zeta_0^2,\sqrt{2\lambda}\zeta)&=&\left(\frac{ 2 j^2(\sqrt{\lambda}\zeta_0)}{\lambda (\zeta^2-\zeta_0^2)}\right)^{1/4}\cos{\mathfrak D},\\
W'(\text{\textonehalf}\lambda\zeta_0^2,\sqrt{2\lambda}\zeta)&=&-\left(\frac{\lambda (\zeta^2-\zeta_0^2)}{ 2 j^{-2}(\sqrt{\lambda}\zeta_0)}\right)^{1/4}\sin{\mathfrak D}, ~~~~\\
W(\text{\textonehalf}\lambda\zeta_0^2,-\sqrt{2\lambda}\zeta)&=&\left(\frac{ 2 j^{-2}(\sqrt{\lambda}\zeta_0)}{\lambda (\zeta^2-\zeta_0^2)}\right)^{1/4} \sin{\mathfrak D},\\
W'(\text{\textonehalf}\lambda\zeta_0^2,-\sqrt{2\lambda}\zeta)&=&-\left(\frac{\lambda (\zeta^2-\zeta_0^2)}{ 2 j^{2}(\sqrt{\lambda}\zeta_0)}\right)^{1/4}\cos{\mathfrak D},
\eqn
with
\bqn
\mathfrak D &\equiv& \frac{1}{2} \lambda \zeta \sqrt{\zeta^2-\zeta_0^2} -\frac{1}{2}\lambda \zeta_0^2 \ln{\left(\frac{\zeta+\sqrt{\zeta^2-\zeta_0^2}}{\zeta_0}\right)}+\frac{\pi}{4}+\phi\left(\frac{1}{2}\lambda\zeta_0^2\right).
\eqn
Here
\bqn
j(x)&=& \sqrt{1+e^{\pi x^2}}-e^{\pi x^2 /2},\\
\phi(x)&=&\frac{x}{2}-\frac{x}{4}\ln{x^2}+\frac{1}{2}\text{ph}\Gamma\left(\frac{1}{2}+ix\right),
\lb{phi}
\eqn
where the phase $\text{ph}\Gamma(\frac{1}{2}+i x)$ is zero when $x=0$ and is determined by continuity otherwise.

\section{Numerical Computation of Particle Production Rate $|\beta_k|^2$}
\renewcommand{\theequation}{C.\arabic{equation}} \setcounter{equation}{0}
\lb{numerical}
In this section, we will outline how to extract the information of particle production rate $|\beta_k|^2$ from the numerical solutions of Eq.({\ref{eom}}). 
The initial condition of the solutions follows Eq.(\ref{initial}) which satisfies the normalization condition, 
\bq
\Big(\phi_1(y),\phi_2(y)\Big)=i(\phi^*_2\phi_1'-\phi_1\phi'^*_2)=-1,
\eq
where $\phi_1(y)$ and $\phi_2(y)$ denote two arbitrary independent solutions of eq.(\ref{eom}). When $y$ approaches zero, the 
wave-function can be expanded in terms of positive and negative frequency modes which solve the asymptotic equation, 
\bq
\frac{d^2\phi_k}{dy^2}+\frac{c}{y^6}\phi_k=0.
\eq
From this equation, one can easily find normalized out-modes, which are,
\bqn
\mu_k(y)&=&\sqrt{\frac{y}{1.27}}\left[J_{-\frac{1}{4}}\left(\frac{\sqrt{c}}{2y^2}\right)-e^{-i \pi/4}J_{\frac{1}{4}}\left(\frac{\sqrt{c}}{2y^2}\right)\right],\\
\nu_k(y)&=&\sqrt{\frac{y}{1.27}}\left[J_{-\frac{1}{4}}\left(\frac{\sqrt{c}}{2y^2}\right)-e^{i \pi/4}J_{\frac{1}{4}}\left(\frac{\sqrt{c}}{2y^2}\right)\right], 
\eqn
where $J_{\nu}(y)$ stands for the Bessel's functions of the first kind and $\mu_k(y)\;\; (\nu_k(y)$) is positive(negative) frequency mode, then $|\beta_k|^2$ can be extracted from the expansion when $y$ is very small,
\bq
\phi_k(y)=\beta_k \mu_k(y)+\alpha_k \nu_k(y), 
\eq
which indicates 
\bq
|\beta_k|^2=\Big( \mu_k(y),\phi_k(y) \Big) \Big( \phi_k(y),\mu_k(y)\Big). 
\eq
Once the numerical solutions of $\phi_k(y)$ is known, one can simply apply the above equation to obtain numerical values of  $|\beta_k|^2$. 

\section{Pair Production in 1+1 dimensional de Sitter space}
\renewcommand{\theequation}{D.\arabic{equation}} \setcounter{equation}{0}

In this Appendix, to show that  {\em the  uniform asymptotic approximation method} used in this paper works well 
for the present problem, we revisit the pair production in 1+1 dimensional de Sitter space, in which the mode 
function can be found exactly and studied in detail   in  \cite{Garriga:1994bm}.  For the case when external electric 
field is turned off, one is required to solve the e.o.m., 
\bq
\lb{d.1}
\phi_k''+\left(k^2+\frac{M^2}{H^2\eta^2}\right)\phi_k=0,
\eq
here $\phi_k$ is the charged scalar field, $M$ denotes its mass, $H$ is the Hubble parameter and the conformal time $\eta$
 takes  the values from the range ($0, +\infty$).  Eq.(\ref{d.1}) has the   solution \cite{Garriga:1994bm},
\bq
\phi_k=\left(\frac{\eta}{8}\right)^{1/2}H^{(2)}_\nu(k\eta).
\eq

On the other hand, in order to use the   uniform asymptotic approximation method, we first notice that this e.o.m. has a second-order pole at origin, thus, we must choose 
\bq
q(\eta)=-\frac{1}{4\eta^2},
\eq
as explained in detail in Section III.E (See also \cite{zhu_constructing_2014,zhu_gravitational_2014,zhu_inflationary_2014,zhu_high-order_2016} for more details.). Therefore, the new e.o.m. takes the form,
\bq
\phi_k''=\left[\lambda^2 \hat g(\eta)+q(\eta)\right]\phi, 
\eq
with
\bq
 \lambda^2 \hat g(\eta)=-k^2+\frac{\nu^2}{\eta^2},~~~~~~\text{and}~~~~~~~\nu=\left(\frac{1}{4}-\frac{M^2}{H^2}\right)^{1/2}.
\eq
Following \cite{zhu_constructing_2014}-\cite{zhu_high-order_2016}, the approximate solution to the first-order of $\lambda^{-1}$  is a linear combination of  the Airy functions ${\mbox{Ai}}(\zeta)$ and ${\mbox{Bi}}(\zeta)$, 
\bq
\label{eomd}
\phi_k=\left(\frac{\zeta(\eta)}{g(\eta)}\right)^{1/4}\Big[\alpha {\mbox{Ai}}(\zeta)+\beta \; {\mbox{Bi}}(\zeta)\Big],
\eq
here $\zeta(\eta)$ is a piecewise function which has the same sign as $\hat g$ in the interval $\eta \in (0, +\infty)$, and is given by
\bqn
\zeta(x) =
\begin{cases}
-\left(\frac{3}{2}\int^x_{x_0}\sqrt{-\hat g}dx\right)^{2/3}, &  \;\; x  >  x_0 = \frac{\nu}{k},\\ 
\left(\frac{3}{2}\int^{x_0}_x\sqrt{g}dx\right)^{2/3}, & \;\;\; 0  < x \le x_0.
\end{cases}
\eqn
Only the asymptotic behaviors of $\zeta$ at $\eta \to +\infty$ and $\eta \to 0^+$ are important,  which are given by
\bqn
\zeta(\eta)&\to& -\left(\frac{3}{2}k\eta-\frac{3\pi\nu}{4}\right)^{2/3}, \;\;\;\; \eta \to +\infty,  \\
\zeta(\eta)&\to&\left(\frac{3}{2}\right)^{2/3}\left(-\nu\ln\eta+\nu\ln\frac{\nu}{k}-\nu+\nu\ln2\right)^{2/3},   \;\;\;\; \eta \to 0^+ .
\eqn
Now we can fix the two free parameters $\alpha$ and $\beta$ appearing  in Eq.(\ref{eomd}) by assuming the Bunch-Davies vacuum at the initial time $\eta\to+\infty$. 
With the help of the asymptotic forms of the Airy functions when its argument approaches negative infinity,
\bqn
Ai(z)&\rightarrow&\frac{1}{\sqrt{\pi}(-z)^{1/4}}\cos\left(\frac{2}{3}(-z)^{3/2}-\frac{\pi}{4}\right),\\
Bi(z)&\rightarrow&-\frac{1}{\sqrt{\pi}(-z)^{1/4}}\sin\left(\frac{2}{3}(-z)^{3/2}-\frac{\pi}{4}\right),
\eqn
we find that 
\bq
\lb{alphabeta}
\alpha=\sqrt{\frac{\pi}{2}},\;\;\;\;
\beta=i\sqrt{\frac{\pi}{2}},
\eq
 so that the initial state is the positive frequency mode,
\bq
\phi_{in}(k)=\frac{1}{\sqrt{2k}}e^{-i\left(k\eta-\frac{\pi\nu}{2}-\frac{\pi}{4}\right)}.
\eq 

\begin{figure}
{\includegraphics[width=8.1cm]{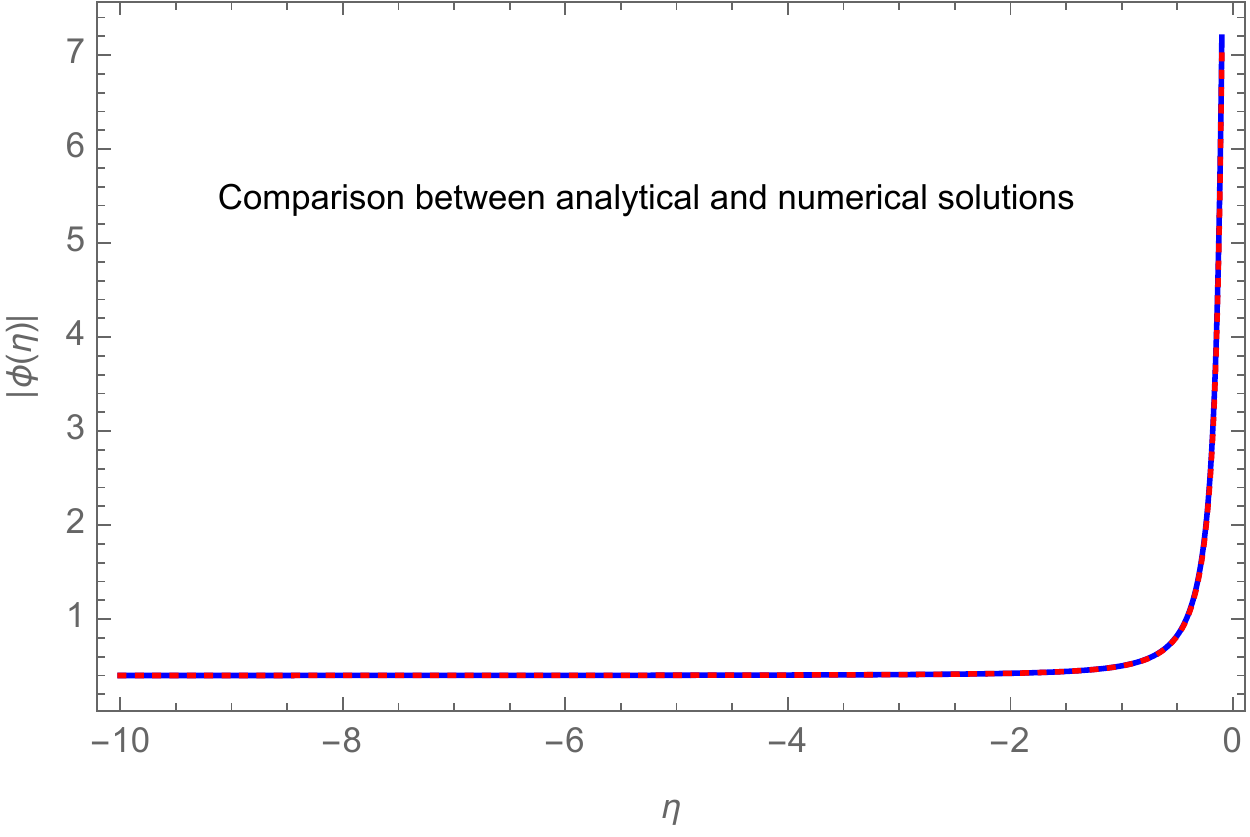}}
\caption{In this figure, the blue dotted curve is from our approximate solution and the   red solid one is the analytical solution presented in \cite{Garriga:1994bm}. In drawing this figure,
we take $k=2$ and $\nu=2$. } \label{2dcase}
\end{figure}

In Fig.\ref{2dcase}, we compare our above approximate solution given by Eq.(\ref{eomd}) with the exact one \cite{Garriga:1994bm}, from which it can be seen 
that the approximate solution traces extremely well to the exact one.

Then,  there will be particles created after time $\nu/k$ for different modes. To show this,   we consider the limit $M\gg H$,  which amounts to taking the parameter $\nu$ to be a pure imaginary 
number $i\epsilon$ with 
\bq
\epsilon\equiv \left(\frac{M^2}{H^2}-\frac{1}{4}\right)^{1/2}.  
\eq
To find particle creation rate at the final time, we need  to know the 
asymptotic forms of the Airy functions at $\eta=0$,  which are given by,
\bqn
\text{Ai} (z)&\rightarrow& \frac{e^{-\frac{2}{3}z^{3/2}}}{2\sqrt{\pi}z^{1/4}},\\
\text{Bi} (ze^{i\pi/3})&\rightarrow&\sqrt{\frac{2}{\pi}}\frac{e^{i\pi/6}}{z^{1/4}}\cos\left(\frac{2}{3}z^{3/2}-\frac{\pi}{4}-\frac{i}{2}\ln2\right).
\eqn
The reason to use this particular form of $\text{Bi} (z)$ is that the phase of $\zeta$ at $\eta \to 0^+$ is larger than $\pi/3$ when we consider 
an imaginary $\nu$. As a result, the final state is a combination of the positive and negative frequency modes 
\bq
\phi_{out}=\left(\frac{\eta}{2\nu}\right)^{1/2}\Bigg[\left(\frac{k\eta}{2}\right)^\nu e^\nu \nu^{-\nu}+\left(\frac{k\eta}{2}\right)^{-\nu}\nu^\nu e^{-\nu-i\pi/2}\Bigg].
\eq
Then,  keeping in mind $\nu=i\epsilon$ and $\eta=-e^{-Ht}/H$, it is straightforward to find
\bq
|\beta_k|^2=\frac{1}{e^{2\pi\epsilon}-1},
\eq
which is  precisely what  found in \cite{Garriga:1994bm}.

\section{Pair Production and WKB condition}
\renewcommand{\theequation}{E.\arabic{equation}} \setcounter{equation}{0}

In this appendix, we compare the particle production effects for the case considered in this paper with $\beta=0$ and the one studied in \cite{Frob:2014zka},
 by considering the WKB analysis for both cases. In general, the solution of the mode function $\phi_k(\eta)$ of the equation,
\bqn
\frac{d^2 \phi_k(\eta)}{d\eta^2} + \Omega_k^2(\eta) \phi_k(\eta)=0,
\eqn 
can be given in terms of the WKB solutions
\bqn
\phi_k(\eta) \simeq \frac{\alpha_k}{\sqrt{2 \Omega_k(\eta)}} e^{- i \int \Omega_k(\eta) d\eta } + \frac{\beta_k}{\sqrt{2\Omega_k(\eta)}} e^{ i \int \Omega_k(\eta) d\eta}, 
\eqn
if the WKB condition
\bqn\lb{WKB}
\left|\frac{3 \Omega_k'^2}{4 \Omega_k^4} - \frac{\Omega_k''}{2 \Omega_k^3}\right| \ll 1,
\eqn
is satisfied. Here we assume that $\Omega_k(\eta)$ is always positive for simplicity.

In order to determine $\alpha_k$ and $\beta_k$ in the above solution, let us consider the WKB adiabatic initial state
\bqn
\phi_k(\eta) = \frac{1}{\sqrt{2\Omega_k(\eta)}} e^{- i \int \Omega_k(\eta) d\eta }.
\eqn
If during the whole process  $ \eta \in(-\infty, 0)$,  the WKB condition (\ref{WKB}) is always satisfied, we will have
\bqn
\alpha_k=1, \;\;\; \beta_k =0.
\eqn
However, in some physical systems, during this process, the WKB condition may be violated or not be fulfilled completely. Then,  the non-adiabatic evolution of the mode $\phi_k(\eta)$ 
will produce excited states (i.e. particle production) during this process and eventually leads to a state with
\bqn
\alpha_k \neq 1, \;\;\; \beta_k \neq 0.
\eqn
The quantity $|\beta_k|^2$ measures the production rate of the excited state.

There are two facts that can lead to the violation of the WKB condition even $\Omega_k^2(\eta)$ is always positive. One case is $\Omega_k(\eta)$ is extreme close to zero and the other is $\Omega_k(\eta)$ is not slowly-varying. 
In the following, we would like to use these facts to discuss the Schwinger particle production effects in the current case   with $\beta=0$ (Sec. 4.0.1) and the case studied in \cite{Frob:2014zka} to show the differences between 
them. In particular, the results obtained in these two cases do not contradict to each other at all, but represent two different physical situations, and both results are correct.

For the case considered in this paper with $\beta=0$, the equation of motion is described by
\bqn
\phi_k''(\eta) + \left[p_a^2+\left(k+\frac{E}{3H^4 \eta^3}\right)^2\right]\phi_k(\eta)=0.
\eqn
Here we have
\bqn
\Omega_k^2(\eta) = p_a^2+\left(k+\frac{E}{3H^4 \eta^3}\right)^2.
\eqn
For $k<0$ it is easy to see $\Omega_k^2(\eta)$ is a monotonic increasing function of $E$. This implies that the particle production rate is suppressed if the electric field becomes larger. For $k>0$, $\Omega_k(\eta)$ has a minimal value at $\eta= - \frac{E^{1/3}}{H^{4/3} k^{1/3}}$. At this point, it is easy to show that the WKB condition now becomes
\bqn
\left|\frac{3 \Omega_k'^2}{4 \Omega_k^4} - \frac{\Omega_k''}{2 \Omega_k^3}\right| =\frac{9H^{8/3} k^{8/3}}{2 E^{2/3} p_a^4}.
\eqn
It is obvious that the WKB condition is   well satisfied  if we increase the value of the electric field $E$. As a result, the particle production is still suppressed if $E$ is large. The results obtained here is in agreement with the quantitative calculation
presented  in Sec. 4.0.1.

For the case in \cite{Frob:2014zka}, the equation of motion reads,
\bqn
\phi_k''(\eta) + \left[\frac{m^2}{H^2 \eta^2}+\left(k-\frac{e E}{H^2 \eta}\right)^2\right]\phi_k(\eta)=0.
\eqn
In this case we have
\bqn
\Omega_k^2(\eta) = \frac{m^2}{H^2 \eta^2}+\left(k-\frac{e E}{H^2 \eta}\right)^2.
\eqn
If $k>0$, then $\Omega_k^2(\eta)$ is a monotonic increasing function of $E$. This implies that the particle production rate is suppressed if the electric field becomes larger. For $k<0$, $\Omega_k(\eta)$ has a minimal value at $\eta= \frac{e E}{H^2 k}$. At this point, it is easy to show that the WKB condition now becomes
\bqn
\left|\frac{3 \Omega_k'^2}{4 \Omega_k^4} - \frac{\Omega_k''}{2 \Omega_k^3}\right| =\frac{e^2 E^2}{2 m^4} + \frac{H^2}{4 m^2}.
\eqn
It is obvious that the WKB condition is not well satisfied if we increase the value of the electric field $E$. This implies that the particle production rate can be enhanced if $E$ is large. The results obtained from the above qualitative analysis are consistent with the ones obtained in \cite{Frob:2014zka}. We can also calculate the particle production effects by using the uniform asymptotic approximation method for this case. In the uniform asymptotic approximation, by choosing 
\bqn
q(y) = - \frac{1}{4 y^2},
\eqn
we have
\bqn
\lambda^2 \hat g(y) = - \frac{1}{y^2}\left(\frac{m^2}{H^2} -\frac{1}{4}\right) - \left(1+ \frac{e E}{H^2 y}\right)^2,
\eqn
where $y=- k \eta$. Then,  $\lambda^2 \hat g(y)=0$ gives two complex conjugated turning points
\bqn
y_1 = \frac{e E}{H^2} - i \sqrt{\frac{m^2}{H^2}-\frac{1}{4}},\nb\\
y_2 = \frac{e E}{H^2} + i \sqrt{\frac{m^2}{H^2}-\frac{1}{4}}.
\eqn
Thus,  from the expression 
\bqn
\lambda \zeta_0^2 = - \frac{2}{\pi} \left| \int_{y_1}^{y_2} \sqrt{\lambda^2 \hat g(y)}dy\right|,
\eqn
we obtain
\bqn
\pi \lambda \zeta_0^2 = - 2 \sqrt{\frac{e^2 E^2}{H^4} + \frac{m^2}{H^2} -  \frac{1}{4}} \mp 2 \frac{e E}{H^2},
\eqn
where $\mp$ corresponding to $k>0$ and $k<0$,  respectively. Thus,  we find
\bqn
|\beta_k|^2=e^{\pi \lambda \zeta_0^2} = e^{- 2 \sqrt{\frac{e^2 E^2}{H^4} + \frac{m^2}{H^2} -  \frac{1}{4}} \mp2 \frac{e E}{H^2}},
\eqn
which is exact the same as that obtained from the exact solution of $\phi_k(\eta)$ in \cite{Frob:2014zka}.


\begin{thebibliography}{99}

\bibitem{Schwinger:1951nm}
  J.~S.~Schwinger,
  {\em On gauge invariance and vacuum polarization},
  Phys.\ Rev.\  {\bf 82}, 664-679 (1951).

\bibitem{Garriga:2012qp} 
  J.~Garriga, S.~Kanno, M.~Sasaki, J.~Soda and A.~Vilenkin,
  {\em Observer dependence of bubble nucleation and Schwinger pair production}, 
 JCAP {\bf 1212}, 006 (2012).
\bibitem{Garriga:2013pga} 
  J.~Garriga, S.~Kanno and T.~Tanaka,
 {\em Rest frame of bubble nucleation}, 
  JCAP {\bf 1306}, 034 (2013)
  [arXiv:1304.6681 [hep-th]].


\bibitem{Garriga:1994bm}
  J.~Garriga,
  {\em Pair production by an electric field in (1+1)-dimensional de Sitter space}, 
  Phys.\ Rev.\  {\bf D49}, 6343-6346 (1994).

  
\bibitem{Kim:2008xv}
  S.~P.~Kim and D.~N.~Page,
  {\em Schwinger Pair Production in dS(2) and AdS(2)}, 
  Phys.\ Rev.\  D {\bf 78}, 103517 (2008)
  [arXiv:0803.2555 [hep-th]].

\bibitem{Kim:2010cb}
  S.~P.~Kim,
  {\em Vacuum Structure of de Sitter Space}, 
    [arXiv:1008.0577 [hep-th]].
    
\bibitem{Frob:2014zka} 
  M.~B.~Frob, J.~Garriga, S.~Kanno, M.~Sasaki, J.~Soda, T.~Tanaka and A.~Vilenkin,
  {\em Schwinger effect in de Sitter space}, 
  JCAP {\bf 1404}, 009 (2014)
  [arXiv:1401.4137 [hep-th]].
  
  

\bibitem{Kobayashi:2014zza} 
  T.~Kobayashi and N.~Afshordi,
 {\em Schwinger Effect in 4D de Sitter Space and Constraints on Magnetogenesis in the Early Universe}, 
  JHEP {\bf 1410}, 166 (2014)
  [arXiv:1408.4141 [hep-th]]; 
  E. Bavarsad, C. Stahl,  and S.-S. Xue, Phys. Rev. D{\bf 94}, 104011 (2016) [arXiv:1507.01686].


  \bibitem{FNPT15} 
  W. Fischler, P.H. Nguyen, J.F. Pedraza, and W. Tangarife, Phys. Rev. D{\bf 91}, 086015 (2015).
  



\bibitem{Hayashinaka:2016qqn} 
  T.~Hayashinaka, T.~Fujita and J.~Yokoyama,
 {\em Fermionic Schwinger effect and induced current in de Sitter space}, 
  JCAP {\bf 1607}, no. 07, 010 (2016)
  [arXiv:1603.04165 [hep-th]];
 E. Bavarsad, C. Stahl,  and S.-S. Xue,   	Phys. Rev. D{\bf 93}, 025004 (2016) [arXiv:1507.01686].
    
\bibitem{Hayashinaka:2016dnt} 
  T.~Hayashinaka and J.~Yokoyama,
  {\em Point splitting renormalization of Schwinger induced current in de Sitter spacetime}, 
  JCAP {\bf 1607}, no. 07, 012 (2016)
  [arXiv:1603.06172 [hep-th]].


\bibitem{SS17} C. Stahl and S.S. Xue, {\em Schwinger effect and backreaction in de Sitter spacetime}, Phys. Lett. B{\bf 760} (2016) 288; 
E. Bavarsad, C. Stahl, and S.S. Xue, {\em Scalar current of created pairs by Schwinger mechanism in de Sitter spacetime}, 
Phys. Rev. D{\bf 94}, 104011 (2016); {\em Fermionic current and Schwinger effect in de Sitter spacetime}, {\em ibid.},  D{\bf 93}, 025004 (2016); 
R. Sharma and S. Singh, {\em Multifaceted Schwinger effects in de Sitter space}, Phys. Rev. D{\bf 96} (2017) 025012 [ arXiv:1704.05076].
  
\bibitem{Watanabe:2009ct}
  M.~a.~Watanabe, S.~Kanno and J.~Soda,
  {\em Inflationary Universe with Anisotropic Hair}, 
  Phys.\ Rev.\ Lett.\  {\bf 102}, 191302 (2009)
  [arXiv:0902.2833 [hep-th]].

\bibitem{Soda:2012zm} 
  J.~Soda,
  {\em Statistical Anisotropy from Anisotropic Inflation}, 
  Class.\ Quant.\ Grav.\  {\bf 29}, 083001 (2012)
  [arXiv:1201.6434 [hep-th]].

\bibitem{Maleknejad:2012fw} 
  A.~Maleknejad, M.~M.~Sheikh-Jabbari and J.~Soda,
 {\em Gauge Fields and Inflation}, 
  Phys.\ Rept.\  {\bf 528}, 161 (2013)
  [arXiv:1212.2921 [hep-th]].

\bibitem{Ratra:1991bn} 
  B.~Ratra,
  {\em Cosmological 'seed' magnetic field from inflation}, 
  Astrophys.\ J.\  {\bf 391}, L1 (1992).

      
\bibitem{Kanno:2009ei}
  S.~Kanno, J.~Soda and M.~a.~Watanabe,
  {\em Cosmological Magnetic Fields from Inflation and Backreaction}, 
  JCAP {\bf 0912}, 009 (2009)
  [arXiv:0908.3509 [astro-ph.CO]].

\bibitem{Emami:2010rm} 
  R.~Emami, H.~Firouzjahi, S.~M.~Sadegh Movahed and M.~Zarei,
 {\em Anisotropic Inflation from Charged Scalar Fields}, 
  JCAP {\bf 1102}, 005 (2011)
  [arXiv:1010.5495 [astro-ph.CO]].



\bibitem{zhu_constructing_2014}
T. Zhu, A. Wang, G. Cleaver, K. Kirsten, and Q. Sheng, 
{\em Constructing analytical solutions of linear perturbations of
  inflation with modified dispersion relations}, 
Int. J. Mod. Phys. A 29, 1450142 (2014).

\bibitem{zhu_gravitational_2014}
T. Zhu, A. Wang, G. Cleaver, K. Kirsten, and Q. Sheng, 
{\em Inflationary cosmology with nonlinear dispersion relations}, 
Phys. Rev. D {\bf 89}, 043507 (2014).

\bibitem{zhu_inflationary_2014}
T. Zhu, A. Wang, G. Cleaver, K. Kirsten, and Q. Sheng,   
{\em Gravitational quantum effects on power spectra and spectral indices with higher-order corrections}, 
Phys. Rev. D {\bf 90}, 063503 (2014).

\bibitem{zhu_high-order_2016}
T. Zhu, A. Wang,  K. Kirsten, G. Cleaver, and Q. Sheng, 
{\em High-order primordial perturbations with quantum gravitational effects}, 
Phys. Rev. D 93, 123525 (2016).

\bibitem{GST} 
A. Gil, J. Segura, and N. M. Temme, {\em Fast and accurate computation of the Weber parabolic cylinder function $W(a,x)$}, IMA J. Numer. Anal. 31, 1194 (2011)

\end{thebibliography}
\end{document}